%% file: main.tex
\documentclass[sigconf]{acmart}
\acmConference[ASE '20]{Melbourne '20: IEEE/ACM International Conference on Automated Software Engineering}{September 21--25, 2020}{Melbourne, AUS}

\usepackage{tikz}
\usepackage{amsmath}

\usepackage{filecontents}

\usepackage[utf8]{inputenc}
\usepackage{markdown}
\usepackage[draft]{minted}
\usepackage{color, colortbl}
\usepackage{censor}
\usepackage{comment}
\usepackage[aboveskip=0em,belowskip=1em]{caption}
\usepackage{draftwatermark}
\SetWatermarkText{Preprint}
\makeatletter
\def\PYG@reset{\let\PYG@it=\relax \let\PYG@bf=\relax%
    \let\PYG@ul=\relax \let\PYG@tc=\relax%
    \let\PYG@bc=\relax \let\PYG@ff=\relax}
\def\PYG@tok#1{\csname PYG@tok@#1\endcsname}
\def\PYG@toks#1+{\ifx\relax#1\empty\else%
    \PYG@tok{#1}\expandafter\PYG@toks\fi}
\def\PYG@do#1{\PYG@bc{\PYG@tc{\PYG@ul{%
    \PYG@it{\PYG@bf{\PYG@ff{#1}}}}}}}
\def\PYG#1#2{\PYG@reset\PYG@toks#1+\relax+\PYG@do{#2}}

\expandafter\def\csname PYG@tok@gd\endcsname{\def\PYG@tc##1{\textcolor[rgb]{0.63,0.00,0.00}{##1}}}
\expandafter\def\csname PYG@tok@gu\endcsname{\let\PYG@bf=\textbf\def\PYG@tc##1{\textcolor[rgb]{0.50,0.00,0.50}{##1}}}
\expandafter\def\csname PYG@tok@gt\endcsname{\def\PYG@tc##1{\textcolor[rgb]{0.00,0.27,0.87}{##1}}}
\expandafter\def\csname PYG@tok@gs\endcsname{\let\PYG@bf=\textbf}
\expandafter\def\csname PYG@tok@gr\endcsname{\def\PYG@tc##1{\textcolor[rgb]{1.00,0.00,0.00}{##1}}}
\expandafter\def\csname PYG@tok@cm\endcsname{\let\PYG@it=\textit\def\PYG@tc##1{\textcolor[rgb]{0.25,0.50,0.50}{##1}}}
\expandafter\def\csname PYG@tok@vg\endcsname{\def\PYG@tc##1{\textcolor[rgb]{0.10,0.09,0.49}{##1}}}
\expandafter\def\csname PYG@tok@vi\endcsname{\def\PYG@tc##1{\textcolor[rgb]{0.10,0.09,0.49}{##1}}}
\expandafter\def\csname PYG@tok@vm\endcsname{\def\PYG@tc##1{\textcolor[rgb]{0.10,0.09,0.49}{##1}}}
\expandafter\def\csname PYG@tok@mh\endcsname{\def\PYG@tc##1{\textcolor[rgb]{0.40,0.40,0.40}{##1}}}
\expandafter\def\csname PYG@tok@cs\endcsname{\let\PYG@it=\textit\def\PYG@tc##1{\textcolor[rgb]{0.25,0.50,0.50}{##1}}}
\expandafter\def\csname PYG@tok@ge\endcsname{\let\PYG@it=\textit}
\expandafter\def\csname PYG@tok@vc\endcsname{\def\PYG@tc##1{\textcolor[rgb]{0.10,0.09,0.49}{##1}}}
\expandafter\def\csname PYG@tok@il\endcsname{\def\PYG@tc##1{\textcolor[rgb]{0.40,0.40,0.40}{##1}}}
\expandafter\def\csname PYG@tok@go\endcsname{\def\PYG@tc##1{\textcolor[rgb]{0.53,0.53,0.53}{##1}}}
\expandafter\def\csname PYG@tok@cp\endcsname{\def\PYG@tc##1{\textcolor[rgb]{0.74,0.48,0.00}{##1}}}
\expandafter\def\csname PYG@tok@gi\endcsname{\def\PYG@tc##1{\textcolor[rgb]{0.00,0.63,0.00}{##1}}}
\expandafter\def\csname PYG@tok@gh\endcsname{\let\PYG@bf=\textbf\def\PYG@tc##1{\textcolor[rgb]{0.00,0.00,0.50}{##1}}}
\expandafter\def\csname PYG@tok@ni\endcsname{\let\PYG@bf=\textbf\def\PYG@tc##1{\textcolor[rgb]{0.60,0.60,0.60}{##1}}}
\expandafter\def\csname PYG@tok@nl\endcsname{\def\PYG@tc##1{\textcolor[rgb]{0.63,0.63,0.00}{##1}}}
\expandafter\def\csname PYG@tok@nn\endcsname{\let\PYG@bf=\textbf\def\PYG@tc##1{\textcolor[rgb]{0.00,0.00,1.00}{##1}}}
\expandafter\def\csname PYG@tok@no\endcsname{\def\PYG@tc##1{\textcolor[rgb]{0.53,0.00,0.00}{##1}}}
\expandafter\def\csname PYG@tok@na\endcsname{\def\PYG@tc##1{\textcolor[rgb]{0.49,0.56,0.16}{##1}}}
\expandafter\def\csname PYG@tok@nb\endcsname{\def\PYG@tc##1{\textcolor[rgb]{0.00,0.50,0.00}{##1}}}
\expandafter\def\csname PYG@tok@nc\endcsname{\let\PYG@bf=\textbf\def\PYG@tc##1{\textcolor[rgb]{0.00,0.00,1.00}{##1}}}
\expandafter\def\csname PYG@tok@nd\endcsname{\def\PYG@tc##1{\textcolor[rgb]{0.67,0.13,1.00}{##1}}}
\expandafter\def\csname PYG@tok@ne\endcsname{\let\PYG@bf=\textbf\def\PYG@tc##1{\textcolor[rgb]{0.82,0.25,0.23}{##1}}}
\expandafter\def\csname PYG@tok@nf\endcsname{\def\PYG@tc##1{\textcolor[rgb]{0.00,0.00,1.00}{##1}}}
\expandafter\def\csname PYG@tok@si\endcsname{\let\PYG@bf=\textbf\def\PYG@tc##1{\textcolor[rgb]{0.73,0.40,0.53}{##1}}}
\expandafter\def\csname PYG@tok@s2\endcsname{\def\PYG@tc##1{\textcolor[rgb]{0.73,0.13,0.13}{##1}}}
\expandafter\def\csname PYG@tok@nt\endcsname{\let\PYG@bf=\textbf\def\PYG@tc##1{\textcolor[rgb]{0.00,0.50,0.00}{##1}}}
\expandafter\def\csname PYG@tok@nv\endcsname{\def\PYG@tc##1{\textcolor[rgb]{0.10,0.09,0.49}{##1}}}
\expandafter\def\csname PYG@tok@s1\endcsname{\def\PYG@tc##1{\textcolor[rgb]{0.73,0.13,0.13}{##1}}}
\expandafter\def\csname PYG@tok@dl\endcsname{\def\PYG@tc##1{\textcolor[rgb]{0.73,0.13,0.13}{##1}}}
\expandafter\def\csname PYG@tok@ch\endcsname{\let\PYG@it=\textit\def\PYG@tc##1{\textcolor[rgb]{0.25,0.50,0.50}{##1}}}
\expandafter\def\csname PYG@tok@m\endcsname{\def\PYG@tc##1{\textcolor[rgb]{0.40,0.40,0.40}{##1}}}
\expandafter\def\csname PYG@tok@gp\endcsname{\let\PYG@bf=\textbf\def\PYG@tc##1{\textcolor[rgb]{0.00,0.00,0.50}{##1}}}
\expandafter\def\csname PYG@tok@sh\endcsname{\def\PYG@tc##1{\textcolor[rgb]{0.73,0.13,0.13}{##1}}}
\expandafter\def\csname PYG@tok@ow\endcsname{\let\PYG@bf=\textbf\def\PYG@tc##1{\textcolor[rgb]{0.67,0.13,1.00}{##1}}}
\expandafter\def\csname PYG@tok@sx\endcsname{\def\PYG@tc##1{\textcolor[rgb]{0.00,0.50,0.00}{##1}}}
\expandafter\def\csname PYG@tok@bp\endcsname{\def\PYG@tc##1{\textcolor[rgb]{0.00,0.50,0.00}{##1}}}
\expandafter\def\csname PYG@tok@c1\endcsname{\let\PYG@it=\textit\def\PYG@tc##1{\textcolor[rgb]{0.25,0.50,0.50}{##1}}}
\expandafter\def\csname PYG@tok@fm\endcsname{\def\PYG@tc##1{\textcolor[rgb]{0.00,0.00,1.00}{##1}}}
\expandafter\def\csname PYG@tok@o\endcsname{\def\PYG@tc##1{\textcolor[rgb]{0.40,0.40,0.40}{##1}}}
\expandafter\def\csname PYG@tok@kc\endcsname{\let\PYG@bf=\textbf\def\PYG@tc##1{\textcolor[rgb]{0.00,0.50,0.00}{##1}}}
\expandafter\def\csname PYG@tok@c\endcsname{\let\PYG@it=\textit\def\PYG@tc##1{\textcolor[rgb]{0.25,0.50,0.50}{##1}}}
\expandafter\def\csname PYG@tok@mf\endcsname{\def\PYG@tc##1{\textcolor[rgb]{0.40,0.40,0.40}{##1}}}
\expandafter\def\csname PYG@tok@err\endcsname{\def\PYG@bc##1{\setlength{\fboxsep}{0pt}\fcolorbox[rgb]{1.00,0.00,0.00}{1,1,1}{\strut ##1}}}
\expandafter\def\csname PYG@tok@mb\endcsname{\def\PYG@tc##1{\textcolor[rgb]{0.40,0.40,0.40}{##1}}}
\expandafter\def\csname PYG@tok@ss\endcsname{\def\PYG@tc##1{\textcolor[rgb]{0.10,0.09,0.49}{##1}}}
\expandafter\def\csname PYG@tok@sr\endcsname{\def\PYG@tc##1{\textcolor[rgb]{0.73,0.40,0.53}{##1}}}
\expandafter\def\csname PYG@tok@mo\endcsname{\def\PYG@tc##1{\textcolor[rgb]{0.40,0.40,0.40}{##1}}}
\expandafter\def\csname PYG@tok@kd\endcsname{\let\PYG@bf=\textbf\def\PYG@tc##1{\textcolor[rgb]{0.00,0.50,0.00}{##1}}}
\expandafter\def\csname PYG@tok@mi\endcsname{\def\PYG@tc##1{\textcolor[rgb]{0.40,0.40,0.40}{##1}}}
\expandafter\def\csname PYG@tok@kn\endcsname{\let\PYG@bf=\textbf\def\PYG@tc##1{\textcolor[rgb]{0.00,0.50,0.00}{##1}}}
\expandafter\def\csname PYG@tok@cpf\endcsname{\let\PYG@it=\textit\def\PYG@tc##1{\textcolor[rgb]{0.25,0.50,0.50}{##1}}}
\expandafter\def\csname PYG@tok@kr\endcsname{\let\PYG@bf=\textbf\def\PYG@tc##1{\textcolor[rgb]{0.00,0.50,0.00}{##1}}}
\expandafter\def\csname PYG@tok@s\endcsname{\def\PYG@tc##1{\textcolor[rgb]{0.73,0.13,0.13}{##1}}}
\expandafter\def\csname PYG@tok@kp\endcsname{\def\PYG@tc##1{\textcolor[rgb]{0.00,0.50,0.00}{##1}}}
\expandafter\def\csname PYG@tok@w\endcsname{\def\PYG@tc##1{\textcolor[rgb]{0.73,0.73,0.73}{##1}}}
\expandafter\def\csname PYG@tok@kt\endcsname{\def\PYG@tc##1{\textcolor[rgb]{0.69,0.00,0.25}{##1}}}
\expandafter\def\csname PYG@tok@sc\endcsname{\def\PYG@tc##1{\textcolor[rgb]{0.73,0.13,0.13}{##1}}}
\expandafter\def\csname PYG@tok@sb\endcsname{\def\PYG@tc##1{\textcolor[rgb]{0.73,0.13,0.13}{##1}}}
\expandafter\def\csname PYG@tok@sa\endcsname{\def\PYG@tc##1{\textcolor[rgb]{0.73,0.13,0.13}{##1}}}
\expandafter\def\csname PYG@tok@k\endcsname{\let\PYG@bf=\textbf\def\PYG@tc##1{\textcolor[rgb]{0.00,0.50,0.00}{##1}}}
\expandafter\def\csname PYG@tok@se\endcsname{\let\PYG@bf=\textbf\def\PYG@tc##1{\textcolor[rgb]{0.73,0.40,0.13}{##1}}}
\expandafter\def\csname PYG@tok@sd\endcsname{\let\PYG@it=\textit\def\PYG@tc##1{\textcolor[rgb]{0.73,0.13,0.13}{##1}}}

\def\PYGZbs{\char`\\}
\def\PYGZus{\char`\_}
\def\PYGZob{\char`\{}
\def\PYGZcb{\char`\}}

\def\PYGZam{\char`\&}
\def\PYGZlt{\char`\<}
\def\PYGZgt{\char`\>}
\def\PYGZsh{\char`\#}

\def\PYGZhy{\char`\-}
\def\PYGZsq{\char`\'}
\def\PYGZdq{\char`\"}

\makeatother

\definecolor{LG}{gray}{0.9}
\definecolor{HG}{gray}{0.6}
\copyrightyear{2020}
\acmYear{2020}
\setcopyright{acmcopyright}\acmConference[ASE '20]{35th IEEE/ACM International Conference on Automated Software Engineering}{September 21--25, 2020}{Virtual Event, Australia}
\acmBooktitle{35th IEEE/ACM International Conference on Automated Software Engineering (ASE '20), September 21--25, 2020, Virtual Event, Australia}
\acmPrice{15.00}
\acmDOI{10.1145/3324884.3416540}
\acmISBN{978-1-4503-6768-4/20/09}
\begin{document}
%don't want date printed
\date{}
%%\SetWatermarkScale{0.5}
%%\StopCensoring
%%\SetWatermarkText{Siemens confidential}
%make title bold and 14 pt font (Latex default is non-bold, 16 pt)
\title{Automated Implementation of Windows-related Security-Configuration Guides}

\author{Patrick Stöckle}
\email{patrick.stoeckle@tum.de}
\affiliation{
    \institution{Technical University of Munich}
    \city{Munich}
    \state{Germany}
}

\author{Bernd Grobauer}
\email{bernd.grobauer@siemens.com}
\affiliation{
    \institution{Siemens AG}
    \city{Munich}
    \state{Germany}
}

\author{Alexander Pretschner}
\email{alexander.pretschner@tum.de}
\affiliation{
    \institution{Technical University of Munich}
    \city{Munich}
    \state{Germany}
}

\input{contents/01_abstract}



\maketitle

\input{contents/02_introduction}
\input{contents/03_generic}

\input{contents/07_windows_automation_poc}
\input{contents/08_evaluation}

\input{contents/09_related_work}
\input{contents/11_conclusion}

\input{contents/12_acknowledgements}
\input{contents/13_availability}

\bibliographystyle{ACM-Reference-Format}

\bibliography{main}
\appendix

\end{document}

%% file: contents/01_abstract.tex
\begin{abstract}
Hardening is the process of configuring IT systems to ensure the security of the systems' components and data they process or store.
The complexity of contemporary IT infrastructures, however, renders manual security hardening and maintenance a daunting task.

In many organizations, security-configuration guides expressed in the \emph{SCAP} (Security Content Automation Protocol) are used as a basis for hardening, but these guides by themselves provide no means for automatically implementing the required configurations.

In this paper, we propose an approach to automatically extract the relevant information from publicly available security-config\-uration guides for Windows operating systems using natural language processing.
In a second step, the extracted information is verified using the information of available settings stored in the \emph{Windows  Administrative Template} files, in which the majority of Windows configuration settings is defined.

We show that our implementation of this approach can extract and implement 83\% of the rules without any manual effort and 96\% with minimal manual effort.
Furthermore, we conduct a study with 12 state-of-the-art guides consisting of 2014 rules with automatic checks and show that our tooling can implement at least 97\% of them correctly. 
We have thus significantly reduced the effort of securing systems based on existing security-configuration guides.

\end{abstract}

%% file: contents/02_introduction.tex
\section{Introduction}
Misconfigurations reduce the security of a system by introducing vulnerabilities that are often difficult to trace.
A recent study~\cite{Dietrich:2018:ISO:3243734.3243794} has demonstrated that from the perspective of the operators there is one major factor for security misconfigurations:
lack of knowledge.

\begin{listing}[b]
\begin{Verbatim}[commandchars=\\\{\}]
\PYG{g+gu}{\PYGZsh{}\PYGZsh{}} /rule
The number of allowed bad logon attempts must be configured to three or less.
\PYG{g+gu}{\PYGZsh{}\PYGZsh{}} /description
The account lockout feature, when enabled, prevents brute\PYGZhy{}force password attacks on the system. The higher this value is, the less effective the account lockout feature will be in protecting the local system. The number of bad logon attempts must be reasonably small to minimize the possibility of a successful password attack while allowing for honest errors made during normal user logon.
\PYG{g+gu}{\PYGZsh{}\PYGZsh{}} /implementations/0/description
Configure the policy value for Computer Configuration \PYGZgt{}\PYGZgt{} Windows Settings \PYGZgt{}\PYGZgt{} Security Settings \PYGZgt{}\PYGZgt{} Account Policies \PYGZgt{}\PYGZgt{} Account Lockout Policy \PYGZgt{}\PYGZgt{} \PYGZdq{}Account lockout threshold\PYGZdq{} to \PYGZdq{}3\PYGZdq{} or fewer invalid logon attempts (excluding \PYGZdq{}0\PYGZdq{}, which is unacceptable).
\end{Verbatim}
\caption{Example of a rule in a Windows-related security-configuration guide.}
\label{lst:rule_example}
\end{listing}
\noindent

One attempt to deal with the lack of knowledge is to use existing security-configuration guides.
These guides consist of several rules for a specific software system such as Windows 10 or Red Hat Enterprise Linux.
Each rule explains which setting should be set to which value to make the system more secure and why we should apply it (e.g., Listing~\ref{lst:rule_example}).
Known publishers of such guides are the Center for Internet Security (CIS) or the Defense Information Systems Agency (DISA).
Organizations and companies like Siemens can use these guides to harden their systems.

One may be tempted to argue that we do not need security configuration because companies like Microsoft make a strong effort to configure their systems securely by default.
These companies invest a lot in security, of course, but security is just one concern, in addition to others, including usability.
Assume that there was a handy setting for most customers, but it poses a small security risk.
The companies may be tempted to decide to have it activated by default, whereas security-aware customers would deactivate it.
Similarly, we could argue for the data collection settings.
They bring knowledge to the companies to improve their products, and all customers can profit from this.
Thus, the companies may be tempted to activate data collection settings by default.
In contrast, customers with high-security requirements would deactivate them to reduce the risk that sensitive information is accidentally leaked via the data collection.
Thus, security-configuration guides from independent organizations can help security-concerned customers in making their systems more secure.

The publishers publish their recommendations on how to configure a software system in formats like PDF and in the Extensible Configuration Checklist Description Format (XCCDF), which is part of the Security Content Automation Protocol (SCAP).
In some cases, these \emph{implementations} are combined with machine-readable and automatable \emph{checks}.
These checks are created manually according to the specification written down in the security-configuration guides.
% BG: some changes here to include information that there
% exist some guides with machine-readable implementation information
% (only for linux).
Although XCCDF is designed as a machine-readable format, instructions for implementing the security settings are only contained in human-readable form in almost all cases.
The notable exception is the OpenSCAP project's~\cite{openscap,preisler:haicman:UsenixLisa:2017} guides for Linux operating systems and applications, which for many rules contain shell scripts and parts of Ansible playbooks.
%BG Below is not really correct: the standard does have the
% fix element (but not more than that, of course)
%The standard does not define how to write \emph{automatable} implementations (c.f., the implementation in Listing~\ref{lst:rule_example}).
Therefore, existing guides solve the lack-of-knowledge problem, but yield another problem:
Automatic \emph{implementations} (or remediation) are not specified in the SCAP standard.
In contrast, the specification of automated \emph{checking} is very detailed.

% BG: Naja, "deal with this problem" finde ich nicht ganz richtig:
% das löst das Problem ja nicht. "Deal with this problm" ist
% eher, dass ab und an extra Mechanismen angeboten werden.
% Ich glaube, darauf muss man eingehen und begründen, warum
% das nicht ausreicht, weil man sich sonst angreifbar macht,
% ein Problem zu lösen, welches gar nicht existiert.
%
% Original Text
% -------------
%Publishers deal with this problem by embedding the actions necessary for the implementation into the remediation description so that administrators easily understand it.
%If we have a system to harden, we automatically check it to get a list of non-compliant rules; for every rule, we manually apply the implementation. Afterward, we automatically recheck the system.
%A frequent approach for hardening a system according to a guide thus is as follows:

Publishers sometimes deal with this problem by providing additional artifacts, such as scripts or -- in the case of Windows -- configuration backup files.
% BG: Let us not call overselves overengineered: we might put thoughts into the head of readers, which were not there before. Also, I think because we have to acknowledge that sometimes there are stand-alone-artifacts provided by publishers for implementation, we have to argue why these are not as helpful as they seem.
%One could now propose that the administrators handcraft a script for every newly published guide to implement it instead of creating an overengineered, automated approach.
The problem here is threefold.
Firstly, such artifacts do not exist for all guides.
Secondly, the guides frequently get updated:
If we take Windows 10 as an example, there will be at least one new guide every year published to deal with the updated settings, e.g., introduced by the version 1909 update;
minor version updates deal with problems or changed requirements.
As a result, DISA, for example, is now at version 18 for its Windows 10 guide.
Therefore, creating/maintaining a mechanism (even if it can be based on some artifact provided by the publisher or)  will be a recurring, manual task.
Thirdly, with stand-alone artifacts for implementation, customization of guides, a feature which is central to SCAP, becomes cumbersome and error-prone, because this requires a manual effort to keep the customized guide in sync with the separately-maintained implementation mechanism.
However, easy customization is essential: Experience shows that there is virtually no use case in which a publicly available security-configuration guide can be implemented without at least some changes.

\begin{figure}[t]
\includegraphics[width=\columnwidth]{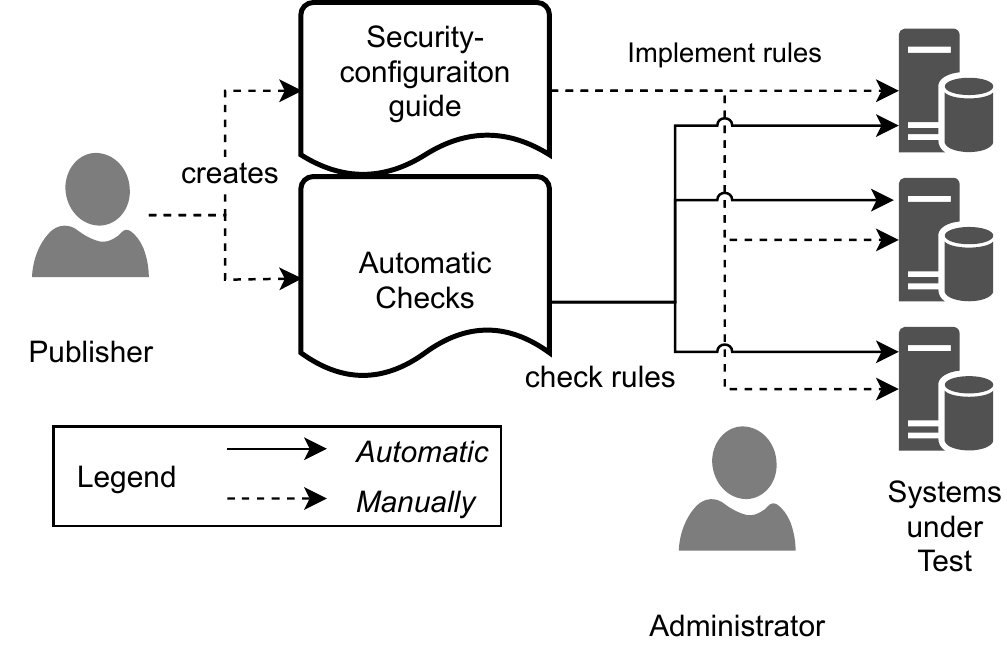}
\caption{Current State of Implementation of Windows-related Security-Configuration Guides}
\label{fig:implementation_process}
\end{figure}

The authoring process is depicted in Figure~\ref{fig:implementation_process}.
The \emph{publisher} creates the guide in the XCCDF format and the corresponding checks in the Open Vulnerability and Assessment Language (OVAL) format.
This is a manual process, as the publishers incorporate their knowledge about the system and its architecture into the guide.
In the next step, an \emph{administrator} uses the automated checks to assess the state of their systems.
The result is a list of the rules to which the system is not compliant;
our evaluation in \S~\ref{sec:evaluation} of over 2000 rules on systems using the default configuration shows that the rate of satisfied rules varies between 0\% and 27\%, with an average of 17.7\%.
Thus, for most of the rules, the (typically: default) configuration of the system to be hardened has to be adjusted.

If the publisher has not provided a mechanism for automated implementation,
for every rule of this list, the administrator must read the implementation/remediation section of the rule in the XCCDF or PDF form of the guide and
implement the steps described there.
If a mechanism is provided, in most cases only a complete implementation of all configuration settings is possible.
This creates significant manual effort for customization, especially if the implementation breaks functionality, but it is unclear which setting(s) have caused the observed problems.

In sum, we address one main \emph{problem}: There are existing guides to configure systems securely, but we cannot implement the required configuration settings (taking into account necessary customization and changes due to updates of the guides) without significant manual effort.

Our \emph{solution} to this problem, realized for Windows operating systems and applications, consists of three major steps.
First, we process the files which define the Windows security policy settings that exist on a Windows-based system.
Windows security policy settings are rules that administrators configure on a computer or multiple devices for the purpose of protecting resources on a device or network.~\cite{secpol}
We can configure a policy setting with a policy path and a value.
The so-called Administrative Template (ADMX/L) files define the majority of policy settings.
They contain information about valid policy paths, possible values for each policy setting, and the underlying implementation of a policy setting within the Windows registry.
Thus, we extract this knowledge in the first step and store it in a machine-readable format to access it during the remediation.
Second, we use natural language processing to extract the settings and the intended values from the guides.
We use the information of the first step to verify that the extracted setting exists and that the extracted value is a valid input for this setting and can, therefore, reduce the risk of wrongly extracted values to a minimum.
Third, we translate the settings and values to their real implementation using the information from the first step.
Our tools can use this information to implement as well as check the configuration settings automatically.

Our contributions are:
\begin{itemize}
    \item an approach to how existing Windows-related security-\\configuration guides can be automatically implemented;
    \item a proof-of-concept implementation of our approach;
    \item a step-by-step documentation of our approach using the DISA Windows Server 2016 guide~\cite{steprepository} and an updated version using the DISA Windows Server 2019~\cite{steprepository2};
    \item an evaluation of our approach using existing guides from DISA and CIS with over 2000 rules~\cite{repoguides}.
\end{itemize}

In \S\ref{sec:genericapproach}, we explain the general idea of our automatic implementation, and in the subchapters, we present the technical details of our proof-of-concept implementation.
In \S\ref{sec:evaluation}, we use the DISA Windows Server 2016 guide and 12 CIS guides to demonstrate the feasibility of our approach.
In \S\ref{sec:generalization}, we discuss challenges and first experiences in generalizing our approach to non-Windows systems as well as additional future work.
\S\ref{sec:related_work} treats related work and \S\ref{sec:conclusion} concludes.

\begin{figure*}
    \includegraphics[width=\textwidth]{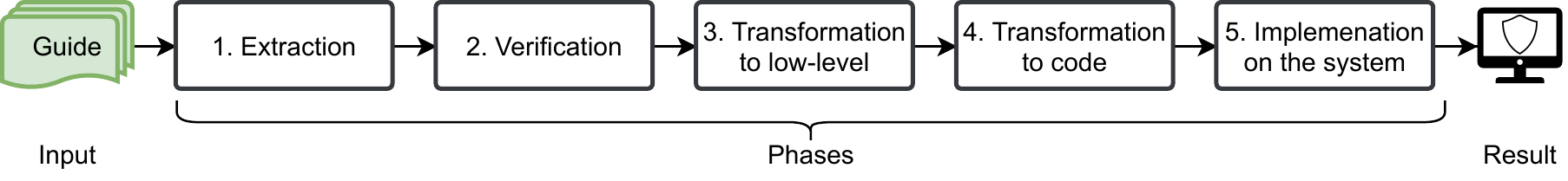}
    \caption{Overview of the abstract hardening approach.}
    \label{fig:hardeningAppraoch}
\end{figure*}

%% file: contents/03_generic.tex
% BG: TODO The tile of this section does not fit, but I really struggle to come up with a better title.
\section{Windows-related Security Configuration}
\label{sec:genericapproach}

\paragraph{Generic Approach}
The generic approach is depicted in Figure~\ref{fig:hardeningAppraoch}.
It shows the different stages of the envisioned process for automatically implementing Windows-related security-configuration guides.
More specifically, the separate steps are defined as follows.

\begin{itemize}
    \item[\textbf{Extraction:}] Use natural language processing (NLP) for each rule to automatically extract the information needed to implement this rule.
    \item[\textbf{Verification:}] Check with an automated mechanism that checks whether the derived information is valid:
    \begin{itemize}
        \item Does the extracted policy path indeed exist?
        \item Has the extracted value the required type for that setting?
        \item Does the extracted value meet the requirements of that setting?
        Is it in the list of possible values or in the range of allowed values?
    \end{itemize}
    If the path or the value is incorrect, the mechanism provides useful feedback about possible paths or values.
    \item[\textbf{Transformation to low-level:}] Transform Windows policy settings into a representation of one of the underlying \textit{low-level} implementation mechanisms.
    This step is necessary because almost none of the most popular configuration-management frameworks can directly process the Windows policy settings, but require the specification of an underlying implementation mechanism:
    \begin{itemize}
        \item Registry settings
        \item Secedit policy file entries
        \item Audit file entries.
    \end{itemize}
    \item[\textbf{Transformation to code:}] Transform these low-level implementation mechanisms into code for carrying out the implementation of each setting.
    \item[\textbf{Implementation:}] Execute code on the system we want to harden to implement the rules.
\end{itemize}
\noindent
We emphasize that especially steps one and two are novel because -- to our best knowledge -- there is no approach published that uses NLP to extract policy settings from SCAP guides, nor is there an approach that verifies extracted values using definition files.
For the evaluation of our approach, we assumed that an evaluation of the complete systems provides more evidence for the usefulness and feasibility of the presented approach than an evaluation of the first two steps alone.
Consequently, we had to design and implement the remaining steps for our PoC implementation.
In the end, we achieved the first published system that reads Windows-related security-configuration guides in the SCAP format and implements them automatically.

%% file: contents/07_windows_automation_poc.tex
%After presenting the general approach, we discuss the %implementation of our approach in detail.
\paragraph{The approach in detail}

We discuss the details of our approach and demonstrate
its feasibility using a proof-of-concept (PoC) implementation.
%BG: wird doch unten nochmalgesagt?
%As example target, we use  a well established and publicly available guide from the DISA for Windows Server 2016 in the SCAP format.
% BG: I do not think the sentence below adds information...
%The use case we considered are organizations that are obliged to harden their systems according to a given guide.

% BG: das wir doch weiter unten nochmal gesagt?
%We executed the different concrete steps of our approach on the given guide, which has been put under version control using git and created a tag after every step.
%Thus, the changes made with a particular step are visible using the diff function of git.
%To ease the understanding, we will provide the links to specific diffs.~\footnote{We use the YAML/Markdown-based \emph{Scapolite} format developed within Siemens, which is better suited than SCAP for authoring and maintenance, to represent the guide. The approach, though, is independent of the YAML-based format.
%and could also be executed directly on the SCAP files.
%}

\begin{figure}[b]
    \includegraphics[width=\columnwidth]{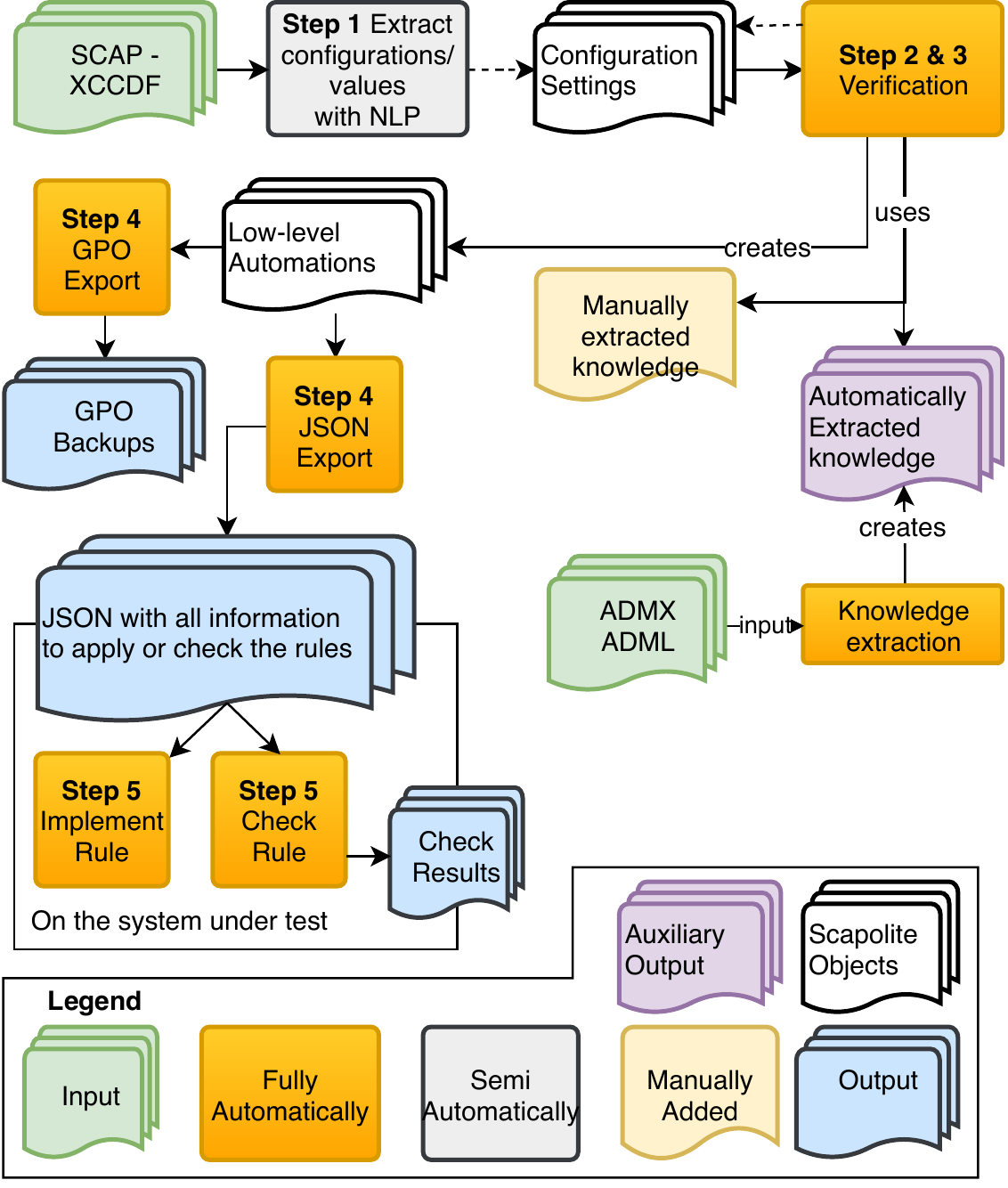}
    \caption{Overview of the steps of our actual implementation.}
    \label{fig:poc}
\end{figure}

The steps of our actual implementation, which we use as a proof of concept, are depicted in Figure~\ref{fig:poc}.
We describe them shortly here and more in detail in the rest of this section.

The input of our PoC consists of guides in the SCAP format.
%A guide consists of rules (for Windows operating systems, several hundreds
%of them), each of which specifies a secure configuration setting.
In the first step, we extract the necessary data for every rule to automate the implementation of this rule using natural language processing.
The result is a set of rules enriched with the configuration settings in a machine-readable format.
These configuration settings are then passed to the verification process:
it has to be verified that the extracted data (a Windows policy path and
required policy values) is valid.
Our implementation uses the information of manually created verification rules
for what essentially are legacy configuration settings combined
with information extracted from the Windows
administrative template files to verify the extracted values.
To make the verification process as fast as possible, we process the latter files a priori and store the information we need in a database format.

If the verification is successful, the low-level automation needed to implement the rule is generated and also stored within the rule.
Depending on the chosen implementation mechanism, these are used to create (1) either a group policy backup, which then can be imported on a Domain Controller to secure all systems in an Active Directory or (2) a JSON file used by a PowerShell script for implementing the settings.
%Alternatively, they can be exported in a JSON format so that our tooling can use these JSON files on the systems under test to implement the rules.
Additionally, our tooling can check the rules using the JSON files, but as SCAP already covers this aspect, we will not look deeper into this facet of our PoC.
\begin{listing}[t]
\begin{Verbatim}[commandchars=\\\{\}]
\PYG{n+nt}{system}\PYG{p}{:} \PYG{l+lScalar+lScalarPlain}{org.scapolite.implementation.win\PYGZus{}gpo}
\PYG{n+nt}{ui\PYGZus{}path}\PYG{p}{:} \PYG{l+lScalar+lScalarPlain}{\PYGZlt{}String containing a valid Windows policy path, using}
            \PYG{l+lScalar+lScalarPlain}{backslashes as separators\PYGZgt{}}
\PYG{n+nt}{value}\PYG{p}{:} \PYG{l+lScalar+lScalarPlain}{\PYGZlt{}A YAML representation of a valid value for the}
        \PYG{l+lScalar+lScalarPlain}{specified path\PYGZgt{}}
\PYG{n+nt}{verification\PYGZus{}status}\PYG{p}{:} \PYG{l+lScalar+lScalarPlain}{(Checked. | Unchecked.)}
\end{Verbatim}
    \caption{Syntax of the Windows policy automation}.
    \label{lst:gpoAutomationA}
\end{listing}

In our PoC implementation, only the second and third steps require a minimum of manual interaction; the other steps are entirely automated.
The dotted line between the \textit{Verification} and the \textit{Configuration Settings} in Figure~\ref{fig:poc} indicates that the person automating the security-configuration guide may have to execute the verification more than once and adjust the values until every rule is marked as \textit{checked} by the verification mechanism.
\begin{figure}[b]
    \includegraphics[width=\columnwidth]{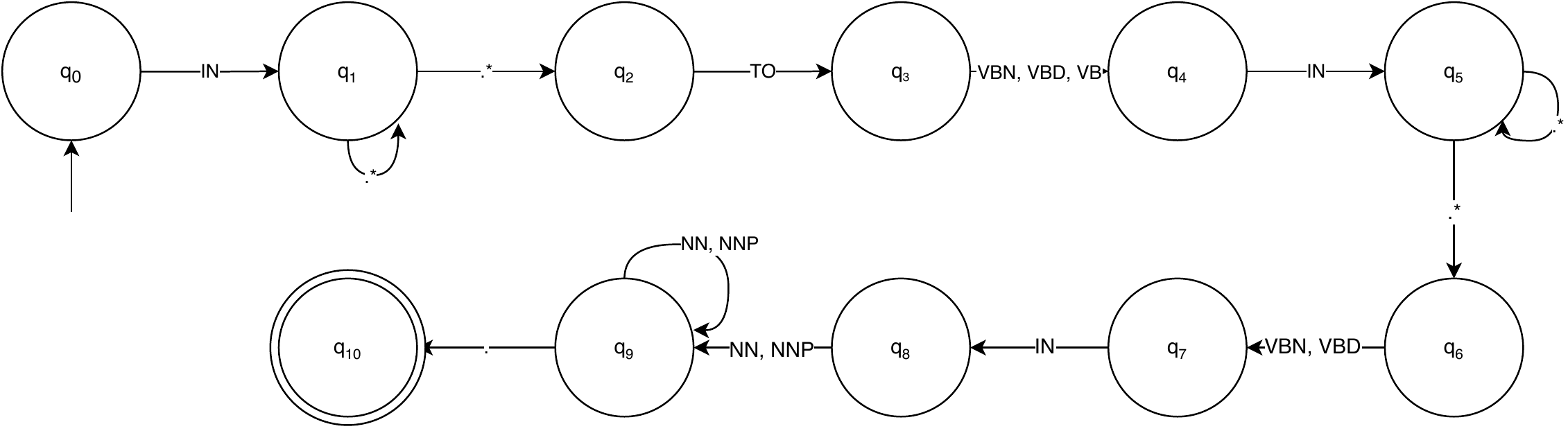}
    \caption{Example of an extraction rule as a nondeterministic finite automaton.}
    \label{fig:stateMachine}
\end{figure}

In the following, we describe each of these steps.
Tooling has been carried out in Python, except for a PowerShell framework for implementing and checking Windows security configurations using the output of Step~4.
As a real-life example, we use the DISA Windows Server 2016 Security Technical Implementation Guide~\cite{iasewindowsserver2016}.
We created a GitHub repository~ \cite{steprepository}, where we conducted all the steps, and created a commit and a tag after every step and reference them by their tags.~\footnote{For representing the guide wihin Github, we use the YAML/Markdown-based \emph{Scapolite} format developed within Siemens, which is better suited than SCAP for authoring and maintenance. The approach, though, is independent of the format.}

\subsection{Natural-language-processing-based extraction of Windows Policy Automations}
\label{sec:nlp_extraction}

The first step of our PoC implementation is the extraction of the needed values using NLP.
Before we can extract the information needed to implement a Windows-related rule automatically, we had to define the structure of the machine-readable constructs, how they are integrated into the rule structure, and what has to be extracted to implement a rule.

For specifying Windows policy settings, the structure must provide information about the \textit{policy path} and the required \textit{value}.
The type of the value (string, list, integer, et cetera) depends on the path;
hence the specification of the automation syntax must refer to the set of valid Windows policy settings as shown in
Listing~\ref{lst:gpoAutomationA}.
Listing~\ref{lst:gpoAutomationB} shows the usage of a policy automation in the rule \emph{SV-88407} of the Windows Server 2016 guide.
\begin{listing}[t]
\begin{Verbatim}[commandchars=\\\{\}]
id: SV\PYGZhy{}88407
rule: \PYGZlt{}see below\PYGZgt{}
implementations:
    \PYG{k}{\PYGZhy{}} description: \PYGZlt{}see below\PYGZgt{}
    automations:
        \PYG{k}{\PYGZhy{}} system: org.scapolite.implementation.win\PYGZus{}gpo
        ui\PYGZus{}path: \PYGZsq{}Computer Configuration\PYGZbs{}Policies\PYGZbs{}Windows Settings\PYGZbs{}Security Settings\PYGZbs{}Local Policies\PYGZbs{}User Rights Assignment\PYGZbs{}Back up files and directories\PYGZsq{}
        value:
        \PYG{k}{\PYGZhy{}} Administrators
\PYGZhy{}\PYGZhy{}\PYGZhy{}

\PYG{g+gu}{\PYGZsh{}\PYGZsh{}} /rule
The Backup files and directories user right must only be assigned to the Administrators group.
\PYG{g+gu}{\PYGZsh{}\PYGZsh{}} /implementations/0/description
Configure the policy value for Computer Configuration \PYGZgt{}\PYGZgt{} Windows Settings \PYGZgt{}\PYGZgt{} Security Settings \PYGZgt{}\PYGZgt{} Local Policies \PYGZgt{}\PYGZgt{} User Rights Assignment \PYGZgt{}\PYGZgt{} \PYGZdq{}Back up files and directories\PYGZdq{} to include only the following accounts or groups:
\PYG{k}{\PYGZhy{}} Administrators
\end{Verbatim}
    \caption{Example rule of the DISA Windows Server 2016 in YAML/Markdown form, incl. a Windows policy automation starting in line 6 (blue).}
    \label{lst:gpoAutomationB}
\end{listing}
\begin{listing}[b]
\begin{Verbatim}[commandchars=\\\{\}]
\PYG{n}{SENTENCE\PYGZus{}WITH\PYGZus{}ENABLED\PYGZus{}WITH\PYGZus{}X\PYGZus{}SELECTED\PYGZus{}FOR\PYGZus{}Y}\PYG{p}{:}
\PYG{p}{\PYGZob{}}\PYG{o}{\PYGZlt{}}\PYG{n}{IN}\PYG{o}{\PYGZgt{}} \PYG{o}{\PYGZlt{}.*\PYGZgt{}+} \PYG{o}{\PYGZlt{}}\PYG{n}{TO}\PYG{o}{\PYGZgt{}} \PYG{o}{\PYGZlt{}}\PYG{n}{VBN}\PYG{o}{|}\PYG{n}{VBD}\PYG{o}{|}\PYG{n}{VB}\PYG{o}{\PYGZgt{}} \PYG{o}{\PYGZlt{}}\PYG{n}{IN}\PYG{o}{\PYGZgt{}} \PYG{o}{\PYGZlt{}.*\PYGZgt{}+} \PYG{o}{\PYGZlt{}}\PYG{n}{VBN}\PYG{o}{|}\PYG{n}{VBD}\PYG{o}{\PYGZgt{}} \PYG{o}{\PYGZlt{}}\PYG{n}{IN}\PYG{o}{\PYGZgt{}} \PYG{o}{\PYGZlt{}}\PYG{n}{NN}\PYG{o}{|}\PYG{n}{NNP}\PYG{o}{\PYGZgt{}+} \PYG{o}{\PYGZlt{}.\PYGZgt{}}\PYG{p}{\PYGZcb{}}
\end{Verbatim}

    \caption{Example of an extraction rule with POS tags.}
    \label{lst:extractionRule}
\end{listing}
In an ideal scenario, the rule already contains the machine-readable automation objects, but this is not the case for guides published in the SCAP.
Thus, we needed to extract the information about required policy settings from the human-readable description in the guide.
To this end, we used the Natural Language Toolkit
(NLTK)~\cite{Bird2009NaturalLanguage}.
Due to the highly schematic structure of the guides under consideration, only eleven extraction rules had to be defined to process most of the rules.
One of the rules is presented here in Listing~\ref{lst:extractionRule} and Figure~\ref{fig:stateMachine}.
The listing shows the definition of such an extraction rule as part of a grammar in NLTK.
\texttt{IN}, \texttt{TO}, etc. refer to the corresponding part-of-speech (POS) tags.
As we have ten rules, our grammar to extract the values consists of ten rule definitions.
To make the idea more precise, Figure~\ref{fig:stateMachine} is presenting the same rule as a nondeterministic finite automaton;
$q_0$ marks the start state and $q_{11}$ the end state.

We use NLTK to label the text of the description of a rule with POS tags.
Afterward, the tagged sentences are passed to the grammar.
If a sentence or a part of a sentence matches an extraction rule, then we know that here we can extract information for the automatic implementation.
We use this sentence from rule \textit{SV-92831} as an example:
\textit{``Configure the policy value for Computer Configuration >> Administrative Templates >> MS Security Guide >> Configure SMBv1 client driver to Enabled with Disable driver (recommended) selected for Configure MrxSmb10 driver.''}
Now, we use NLTK to get the POS tags:
\textit{('Configure', 'VB'), ('the', 'DT'), ..., ('for', 'IN'), ('Computer', 'NNP'), ..., ('driver', 'NN'), ('to', 'TO'), ('Enabled', 'VB'), ('with', 'IN'), ('Disable', 'JJ'), ...,  (')', ')'), ('selected', 'VBN'), ('for', 'IN'), ('Configure', 'NNP'), ..., ('driver', 'NN'), ('.', '.')}
The segment starting at \textit{for} matches the pattern defined in the extraction rule, and we would reach the end state of Figure~\ref{fig:stateMachine}.
Using our definition of the extraction rule, we know that we have the policy path in the part within the POS tags \textit{IN} and \textit{TO}, the first value between \textit{TO} and \textit{IN}, the second value between \textit{IN} and \textit{VBN, IN}, and the name of the option for which the second value has to be set between \textit{VBN, IN}, and \text{'.'}.

As already mentioned, we need only eleven extraction rules to extract information for most of the DISA Windows Server 2016 guide;
for a comparable CIS guide, we defined ten rules.
Please note that the extraction using NLP is as simple as this only because DISA and CIS write their guides in a highly schematic way.

If the automatic extraction process could not obtain any or only ambiguous information for a setting to set, the respective rules are marked in this step of the process.
For these rules, automation objects have to be created manually using the hints from the automatic extraction.
For the analysis of the degree of automation, we refer to \S\ref{sub:degreeofautomation}.
Listing~\ref{lst:gpoAutomationB} is the result of a successful extraction carried out by our tool.\footnote{Tag: \href{https://github.com/tum-i22/disa-windows-server-2016/releases/tag/step-3-extract-configurations-values-with-nlp}{step-3-extract-configurations-values-with-nlp}}

\subsection{Verification of Windows policy automations}
\label{sec:gpo_verification}

As already mentioned in \S\ref{sec:nlp_extraction}, the set of available policy settings determine the syntax (and semantics) of the Windows policy automations.
The set of available policy settings varies between different versions of operating systems and policy-managed applications.
Thus, we can determine the validity of a policy automation for a specific version of OS or an application.
\begin{listing}[t]
\begin{Verbatim}[commandchars=\\\{\}]
\PYG{n+nt}{id}\PYG{p}{:} \PYG{l+lScalar+lScalarPlain}{controlpaneldisplay\PYGZus{}\PYGZus{}cpl\PYGZus{}personalization\PYGZus{}nolockscreencamera}
\PYG{n+nt}{registry}\PYG{p}{:}
    \PYG{n+nt}{name}\PYG{p}{:} \PYG{l+lScalar+lScalarPlain}{NoLockScreenCamera}
    \PYG{n+nt}{path}\PYG{p}{:} \PYG{l+lScalar+lScalarPlain}{Software\PYGZbs{}Policies\PYGZbs{}Microsoft\PYGZbs{}Windows\PYGZbs{}Personalization}
    \PYG{n+nt}{hive}\PYG{p}{:} \PYG{l+lScalar+lScalarPlain}{HKEY\PYGZus{}LOCAL\PYGZus{}MACHINE}
    \PYG{n+nt}{type}\PYG{p}{:} \PYG{l+lScalar+lScalarPlain}{REG\PYGZus{}DWORD}
    \PYG{n+nt}{enabled\PYGZus{}value}\PYG{p}{:} \PYG{l+lScalar+lScalarPlain}{1}
    \PYG{n+nt}{disabled\PYGZus{}value}\PYG{p}{:} \PYG{l+lScalar+lScalarPlain}{0}
\end{Verbatim}
    \caption{Example of a relationship between the id and the definition of the registry to set. }
    \label{lst:idToRegistry}
\end{listing}
\noindent
As mentioned before, the ADMX/L files define the majority of Windows policy settings.
The Windows OSs use these files to display the GUI for configuring policy settings via point-and-click and keep the policy content and the actual implementation of the settings in the registry in sync.
Microsoft regularly issues updates of the ADMX/L files.

%We have written an importer for ADMX/L files that is to be run whenever new ADMX/L files are published; the relevant information is extracted and stored for usage by our toolset in this validation step as well as the subsequent step (\S\ref{sec:generate_low_level_mechanisms}).

To make this more visual, we provide another example:
From the \textit{ControlPanelDisplay.admx} and the \textit{ControlPanelDisplay.adml} files located under policies on Windows Server 2016 instances, our exporter can get the information that the setting with the policy path \textit{Computer Configuration \textbackslash{} ... Control Panel \textbackslash{} Personalization \textbackslash{} Prevent enabling lock screen camera } has the id \textit{CPL\_ Personalization\_ NoLockScreenCamera}.
We store this relationship and the information to which registry this id belongs in our export;
this is presented in Listing~\ref{lst:idToRegistry}.
Here, we can get the information on which hive, path, and registry name are affected.
Furthermore, we know that only \textit{Enabled} and \textit{Disabled} are valid options for this setting and that we can translate them to 1 and 0, respectively.
%Our importer is based on the implementation of the \verb+win_lgpo+ of SaltStack~\cite{saltstack};
%we parallelized the code to make it faster and adjusted it to store the information so that we do not have to construct these relationships on the fly.

There are, however, also Windows policy settings that are not defined via ADMX/L files.
These other settings are represented through entries in either a special configuration file (\texttt{GptTmpl.INF}) or a CSV file (\texttt{audit.CSV}) when creating a file-based representation of policy settings on a Windows OS through the \verb+lgpo.exe+~\cite{lgpo} tool provided by Microsoft.
Unfortunately, there exists -- to our best knowledge -- no machine-readable representation that specifies these policy settings.
Luckily, we could extract many of these specifications for configuration definitions from the SaltStack~\cite{saltstack} implementation of the \verb+win_lgpo+ module for managing Windows configuration settings. (From the 196 settings configurable via the INF file, we could obtain 139 from SaltStack's implementation; the remaining specifications, which we encountered in the course of our work on several Windows OS versions, were added manually.)
Furthermore, we could extract the specifications for all settings handled via \texttt{audit.CSV} via parsing a given \texttt{audit.CSV} file.
Thus, the manual effort required for dealing with these non-ADMX/L settings was negligible when compared to the over 4000 configuration specifications we could extract automatically.

With the information of the knowledge extraction, the verification process can now determine for each configuration setting if the policy path is valid and, if so, whether the provided value is admissible for that particular policy path.

We have implemented our tooling such that the Windows policy automations in a given guide are parsed and verified.
If the policy path exists and the given value is acceptable, the automation is marked as checked.
If not, the automation is enriched with as much information as possible:
\begin{itemize}
    \item If the policy path does not exist, information about similar policy paths is supplied, using the Levenshtein distance~\cite{levenshtein1966binary} on character and word basis over the set of valid policy paths.
    This set is a byproduct of our import step.
    To have the set of valid policy paths accessible is one reason to create those files a priori.
    Listing~\ref{lst:verification_wrong_value_example} a) provides an example of the result of the verification step.
    \item If the value is not admissible for the given policy path, information about admissible values is added to the automation -- see Listing~\ref{lst:verification_wrong_value_example}b) and c).
\end{itemize}

\begin{listing}[!htb]
\begin{Verbatim}[commandchars=\\\{\}]
\PYG{n+nt}{ui\PYGZus{}path}\PYG{p}{:} \PYG{l+lScalar+lScalarPlain}{... \PYGZbs{} Control Panel \PYGZbs{} Personalization \PYGZbs{} Prevent}
    \PYG{l+lScalar+lScalarPlain}{enabling lock screen}
\PYG{n+nt}{value}\PYG{p}{:} \PYG{l+lScalar+lScalarPlain}{Enabled}
\PYG{n+nt}{error\PYGZus{}class}\PYG{p}{:} \PYG{l+lScalar+lScalarPlain}{NOT\PYGZus{}FOUND policy name \PYGZdq{}preventenablinglockscreen\PYGZdq{}}
\PYG{n+nt}{error\PYGZus{}hint}\PYG{p}{:} \PYG{l+s}{\PYGZdq{}}\PYG{n+nv}{ }\PYG{l+s}{The}\PYG{n+nv}{ }\PYG{l+s}{given}\PYG{n+nv}{ }\PYG{l+s}{path}\PYG{n+nv}{ }\PYG{l+s}{was}\PYG{n+nv}{ }\PYG{l+s}{not}\PYG{n+nv}{ }\PYG{l+s}{found,}\PYG{n+nv}{ }\PYG{l+s}{but}\PYG{n+nv}{ }\PYG{l+s}{there}\PYG{n+nv}{ }\PYG{l+s}{were}\PYG{n+nv}{ }\PYG{l+s}{3}\PYG{n+nv}{ }\PYG{l+s}{similar}\PYG{n+nv}{ }\PYG{l+s}{policies.}\PYG{n+nv}{ }\PYG{l+s}{If}\PYG{n+nv}{ }\PYG{l+s}{the}\PYG{n+nv}{ }\PYG{l+s}{UI}\PYG{n+nv}{ }\PYG{l+s}{path}\PYG{n+nv}{ }\PYG{l+s}{you}\PYG{n+nv}{ }\PYG{l+s}{were}\PYG{n+nv}{ }\PYG{l+s}{looking}\PYG{n+nv}{ }\PYG{l+s}{for}\PYG{n+nv}{ }\PYG{l+s}{is}\PYG{n+nv}{ }\PYG{l+s}{in}\PYG{n+nv}{ }\PYG{l+s}{the}\PYG{n+nv}{ }\PYG{l+s}{array,}\PYG{n+nv}{ }\PYG{l+s}{please}\PYG{n+nv}{ }\PYG{l+s}{replace}\PYG{n+nv}{ }\PYG{l+s}{the}\PYG{n+nv}{ }\PYG{l+s}{original}\PYG{n+nv}{ }\PYG{l+s}{UI}\PYG{n+nv}{ }\PYG{l+s}{path}\PYG{n+nv}{  }\PYG{l+s}{with}\PYG{n+nv}{ }\PYG{l+s}{the}\PYG{n+nv}{ }\PYG{l+s}{new}\PYG{n+nv}{ }\PYG{l+s}{UI}\PYG{n+nv}{ }\PYG{l+s}{path.\PYGZdq{}}
\PYG{n+nt}{candidates}\PYG{p}{:}
\PYG{p+pIndicator}{\PYGZhy{}} \PYG{l+lScalar+lScalarPlain}{Control Panel\PYGZbs{}Personalization\PYGZbs{}Prevent enabling lock screen camera}
\PYG{p+pIndicator}{\PYGZhy{}} \PYG{l+lScalar+lScalarPlain}{... \PYGZbs{} Prevent enabling lock screen slide show}
\PYG{p+pIndicator}{\PYGZhy{}} \PYG{l+lScalar+lScalarPlain}{... \PYGZbs{} Prevent changing the color scheme}
\PYG{n+nn}{\PYGZhy{}\PYGZhy{}\PYGZhy{}}
\PYG{n+nt}{ui\PYGZus{}path}\PYG{p}{:} \PYG{l+s}{\PYGZsq{}...}\PYG{n+nv}{ }\PYG{l+s}{\PYGZbs{}}\PYG{n+nv}{ }\PYG{l+s}{Network}\PYG{n+nv}{ }\PYG{l+s}{security:}\PYG{n+nv}{ }\PYG{l+s}{LAN}\PYG{n+nv}{ }\PYG{l+s}{Manager}\PYG{n+nv}{ }\PYG{l+s}{authentication}\PYG{n+nv}{ }\PYG{l+s}{level\PYGZsq{}}
\PYG{n+nt}{value}\PYG{p}{:} \PYG{l+lScalar+lScalarPlain}{Send NTLMv2 response}
\PYG{n+nt}{error\PYGZus{}class}\PYG{p}{:} \PYG{l+lScalar+lScalarPlain}{CONFIGURE}
\PYG{n+nt}{error\PYGZus{}hint}\PYG{p}{:} \PYG{l+s}{\PYGZdq{}To}\PYG{n+nv}{ }\PYG{l+s}{apply}\PYG{n+nv}{ }\PYG{l+s}{this}\PYG{n+nv}{ }\PYG{l+s}{rule,}\PYG{n+nv}{ }\PYG{l+s}{please}\PYG{n+nv}{ }\PYG{l+s}{choose}\PYG{n+nv}{ }\PYG{l+s}{a}\PYG{n+nv}{ }\PYG{l+s}{setting}\PYG{n+nv}{ }\PYG{l+s}{value}\PYG{n+nv}{ }\PYG{l+s}{for}\PYG{n+nv}{ }\PYG{l+s}{each}\PYG{n+nv}{ }\PYG{l+s}{sub\PYGZhy{}setting}\PYG{n+nv}{ }\PYG{l+s}{in}\PYG{n+nv}{ }\PYG{l+s}{candidates.}\PYG{n+nv}{ }\PYG{l+s}{Next,}\PYG{n+nv}{ }\PYG{l+s}{replace}\PYG{n+nv}{ }\PYG{l+s}{the}\PYG{n+nv}{ }\PYG{l+s}{content}\PYG{n+nv}{ }\PYG{l+s}{of}\PYG{n+nv}{ }\PYG{l+s}{the}\PYG{n+nv}{ }\PYG{l+s}{\PYGZsq{}value\PYGZsq{}}\PYG{n+nv}{ }\PYG{l+s}{attribute}\PYG{n+nv}{ }\PYG{l+s}{with}\PYG{n+nv}{ }\PYG{l+s}{the}\PYG{n+nv}{ }\PYG{l+s}{content}\PYG{n+nv}{ }\PYG{l+s}{of}\PYG{n+nv}{ }\PYG{l+s}{candidates.\PYGZdq{}}
\PYG{n+nt}{candidates}\PYG{p}{:}
\PYG{p+pIndicator}{\PYGZhy{}} \PYG{l+lScalar+lScalarPlain}{Send LM \PYGZam{} TLM responses \PYGZhy{} use NTLMv2 session security if negotiated}
\PYG{p+pIndicator}{\PYGZhy{}} \PYG{l+lScalar+lScalarPlain}{Send NTLMv2 response only. Refuse LM \PYGZam{} NTLM}
\PYG{p+pIndicator}{\PYGZhy{}} \PYG{l+lScalar+lScalarPlain}{Send NTLM response only}
\PYG{n+nn}{...}
\PYG{n+nn}{\PYGZhy{}\PYGZhy{}\PYGZhy{}}
\PYG{n+nt}{ui\PYGZus{}path}\PYG{p}{:} \PYG{l+lScalar+lScalarPlain}{... \PYGZbs{} Configure Windows Defender SmartScreen}
\PYG{n+nt}{value}\PYG{p}{:} \PYG{l+lScalar+lScalarPlain}{Enabled}
\PYG{n+nt}{candidates}\PYG{p}{:}
    \PYG{n+nt}{main\PYGZus{}setting}\PYG{p}{:}
        \PYG{p+pIndicator}{\PYGZhy{}} \PYG{l+lScalar+lScalarPlain}{Disabled}
        \PYG{p+pIndicator}{\PYGZhy{}} \PYG{l+lScalar+lScalarPlain}{Enabled}
    \PYG{n+nt}{Pick one of the following settings}\PYG{p}{:}
        \PYG{p+pIndicator}{\PYGZhy{}} \PYG{l+lScalar+lScalarPlain}{Warn}
        \PYG{p+pIndicator}{\PYGZhy{}} \PYG{l+lScalar+lScalarPlain}{Disabled}
        \PYG{p+pIndicator}{\PYGZhy{}} \PYG{l+lScalar+lScalarPlain}{Warn and prevent bypass}
\end{Verbatim}

\caption{Failed verifications:
a)
Policy path does not exist; information about 3 possible options.
b)
Specified value does not exist; admissible values provided.
c) Policy setting underspecified; request for additional value.
}
\label{lst:verification_wrong_value_example}
\end{listing}
\noindent
We proceed as follows to verify and correct the policy automations:
\begin{enumerate}
    \item The verification mechanism is run a first time.\footnote{Tag: \href{https://github.com/tum-i22/disa-windows-server-2016/releases/tag/step-4-verification-1}{step-4-verification-1}}
    \item The user reviews the reported errors and corrects them.
    \item Verification is re-run either on a rule-by-rule basis or for the complete guide.\footnote{Tag: \href{https://github.com/tum-i22/disa-windows-server-2016/releases/tag/step-4c-fix}{step-4c-fix}}
    \item Once all errors have been corrected, an export pairing the human-readable description and the policy automation for each rule is created, allowing the user to verify very quickly that the automation indeed faithfully reflects the human-readable specification.\footnote{Tag: \href{https://github.com/tum-i22/disa-windows-server-2016/releases/tag/step-5-create-xlsx-report-for-the-current-guide}{step-5-create-xlsx-report-for-the-current-guide}}
\end{enumerate}
\noindent
This verification seems simple, but studies have shown that 42\% of the configuration errors that caused high-impact incidents are obvious errors (e.g., typos) \cite{Tang:2015:HCM:2815400.2815401} and that
a significant number of configuration errors are due to compatibility issues\cite{Yin:2011:ESC:2043556.2043572}.
Our verification is able to catch such problems at the earliest possible stage.
%%BG: open etwas umformuliert.
% Der Paragraph endete mit "would reduce the number of configuration errors dramatically if we removed these errors. "

\subsection{Generation of low-level implementation mechanisms}
\label{sec:generate_low_level_mechanisms}
Windows policy settings are implemented through registry settings, \textit{INF} policy file entries, and audit file entries.
To represent these mechanisms within a guide, we introduce automation extensions for these three mechanisms.
Using the information gathered as described in \S\ref{sec:gpo_verification}, we implemented a transformation from the policy automation into the corresponding \textit{low-level} automation extension.\footnote{Tag: \href{https://github.com/tum-i22/disa-windows-server-2016/releases/tag/step-6-enrich-scapolite-with-low-level-automations}{step-6-enrich-scapolite-with-low-level-automations}, a table with all the low-level automations can be found under \href{https://github.com/tum-i22/disa-windows-server-2016/blob/master/xlsx/report_with_low_level_automations.xlsx}{xlsx/report\_with\_low\_level\_automations.xlsx}. }

Listing~\ref{lst:registry_automation_example} provides an example:
according to the Windows policy automation (line 1 to 4), the value \textit{Enabled} has to be set for the policy setting with path \textit{... \textbackslash{} Apply UAC restrictions to local accounts on network logons}.
Using information extracted from the ADMX/L files, we can generate the Windows registry automation:
the registry key under the path {\small \textit{SOFTWARE \textbackslash{} Microsoft \textbackslash{} Windows\textbackslash{}CurrentVersion\textbackslash{}Policies\textbackslash{}System} } with the value name \textit{LocalAccountTokenFilterPolicy} has to be set to a \textit{DWORD} with the value 0.
\begin{listing}[t]
\begin{Verbatim}[commandchars=\\\{\}]
\PYG{n+nt}{ui\PYGZus{}path}\PYG{p}{:} \PYG{l+lScalar+lScalarPlain}{...\PYGZbs{}Apply UAC restrictions to local accounts on network logons}
\PYG{n+nt}{value}\PYG{p}{:} \PYG{l+lScalar+lScalarPlain}{Enabled}
\PYG{n+nt}{verification\PYGZus{}status}\PYG{p}{:} \PYG{l+lScalar+lScalarPlain}{Checked.}
\PYG{p+pIndicator}{\PYGZhy{}} \PYG{n+nt}{system}\PYG{p}{:} \PYG{l+lScalar+lScalarPlain}{org.scapolite.implementation.windows\PYGZus{}registry}
\PYG{n+nt}{config}\PYG{p}{:} \PYG{l+lScalar+lScalarPlain}{Computer}
\PYG{n+nt}{registry\PYGZus{}key}\PYG{p}{:} \PYG{l+lScalar+lScalarPlain}{SOFTWARE\PYGZbs{}Microsoft\PYGZbs{}Windows\PYGZbs{}CurrentVersion\PYGZbs{}Policies\PYGZbs{}System}
\PYG{n+nt}{value\PYGZus{}name}\PYG{p}{:} \PYG{l+lScalar+lScalarPlain}{LocalAccountTokenFilterPolicy}
\PYG{n+nt}{action}\PYG{p}{:} \PYG{l+lScalar+lScalarPlain}{DWORD:0}
\end{Verbatim}
    \caption{Example of a Windows policy automation and the resulting Windows registry automation.}
    \label{lst:registry_automation_example}
\end{listing}

\subsection{Transformation into code}
\label{section:generate_code}

The main idea between the separation of this step and the actual implementation was that we could execute all the previous steps on one machine, export the information, and do only the actual implementation on the system under test.
%BG: Ist Untenstehendes nicht
% offensichtlich?
%Thus, only Python has to be installed, and also the files needed for the verification are only present on that machine.
% BG: unten etwas umformuliert
Thus, a central instance can be used for storing and processing relevant guides; systems under test can fetch the required data for implementing (and testing) security configurations from that central server.
To further facilitate this approach, we implemented an export from a guide containing \textit{low-level} automation for Windows into a JSON document that contains all data relevant for implementing each rule with the associated automation(s).

In order to support implementations via policies (either via local group policies or via the Active Directory capabilities), we can also automatically generate policy backups based on the extracted information.
We have implemented this step as part of a continuous-integration where changes to automation in a guide lead to an automated re-generation of both scripts and policy backups.\footnote{Tag: \href{https://github.com/tum-i22/disa-windows-server-2016/releases/tag/step-8-export}{step-8-export}, policy backup folder for each profile under \href{https://github.com/tum-i22/disa-windows-server-2016/tree/step-8-export/lgpo_backups}{lgpo\_backups}.}

\subsection{Implementation of the rules on the system using PowerShell}

When choosing a target language framework to use to implement the rules using the information of the tow-level automation described in \S\S\ref{sec:gpo_verification}~and~\ref{sec:generate_low_level_mechanisms}, we decided to use PowerShell for the following reasons:

\begin{itemize}
    \item Common configuration management frameworks like Ansible, Puppet, Chef, and SaltStack cannot handle the Windows policy settings or use PowerShell to implement them.
    Thus, we decided to use PowerShell without a configuration framework as a wrapper to implement the rules.
    \item Microsoft's efforts to allow code/script-based configuration management of Windows rather than the GUI-based mechanism centering on the policy editor are based on PowerShell.
    \item PowerShell is installed by default on all Windows OSs that are still in mainstream support by Microsoft.
    \item To fully leverage the ability to generate mechanisms for rule-by-rule implementation rather than the \textit{bulk} implementation offered, e.g., in the form of policy backups, we looked for robust roll-back functionality that allowed us to reset a configuration reliably to its previous value.
\end{itemize}

Thus, we have created a PowerShell library that, based on the JSON file, applies, checks, and reverts single as well as several or all rules.
As mentioned before, our tooling uses the extracted information to check whether the system is compliant to a rule automatically.
This functionality is already covered within SCAP, and there are many SCAP-compliant scanners.
Therefore, the checking functionality is not in the focus of this paper.

Our PowerShell library uses Windows tools that assure that the configuration changes are reflected in the local policy: \textit{secedit}, \textit{auditpol}, and \textit{LGPO.exe}~\cite{lgpo}.
In the end, we can implement a security-configuration guide by running one PowerShell command.

%% file: contents/08_evaluation.tex
\section{Evaluation}
\label{sec:evaluation}

\begin{table}[t]
    \caption{Extracted, verified, and automated rules.}
    \label{table:automation}
    \begin{tabular}{l r r r}
        \toprule
        Categories                              & \#    & \%    & \% of OVAL \\
        \midrule
        Rules                                   & 274   & 100   & \\ \midrule
        Configurations Extracted with NLP       & 198   & 72.3  & 95.6    \\
        Rules without extracted values          & 76    & 27.7  & 36.7 \\  \midrule
        First-Time Verified                     & 173   & 63.1  & 83.6    \\
        Not verified the first time             & 25    & 9.1   & 12.1    \\  \midrule
        Non-automatable but extracted           & 2     & 0.7   & 1.0   \\
        Automatable but not extracted           & 4     & 1.5   & 1.9    \\  \midrule
        Verified after manual correction        & 27    & 9.9   & 13.0    \\
        Automated Rules                         & 200   & 73.0  & 96.6    \\
      \bottomrule
    \end{tabular}
\end{table}

To demonstrate the presented approach's potential, we use the real-life example of realizing automatic rule-by-rule implementations for the DISA Microsoft Windows Server 2016 guide Benchmark~\cite{iasewindowsserver2016} for an evaluation.\footnote{
We choose DISA's guide because their SCAP content is public.
Only CIS members can access CIS's SCAP content, whereas their PDFs are publicly available.}
The benchmark consists of 207 rules with automatic checks and 67 rules without automatic checks.

The results of all steps shown below are available for review \cite{steprepository}.
Every step is denoted as a commit and marked with a tag.
Thus, a diff view between a commit and its predecessor reveals the constructs added, removed, or changed in this step.
In this article, we will concentrate on this repository.
Additionally, we created a new repository with the DISA Windows Server 2019 guide\cite{steprepository2} and executed the same steps to demonstrate that our approach works on recent SCAP documents as well.
Thus, the fact that we used Windows Server 2016 should not be a threat to our evaluation's validity.

We seek to answer the following \textbf{Research Questions}:
\begin{description}
    \item[RQ1] For how many rules can we automatically derive an implementation from the text in natural language?
    How high is their percentage?
    \item[RQ2] How many of the extracted rules are automatable, and how many automatable rules were not extracted?
    \item[RQ3] After correcting wrongly extracted automations:
    How many rules can we implement automatically for the complete guide?
    \item[RQ4] How much time does our approach require to extract the information, verify it, and implement the rule?
    \item[RQ5] How many rules are implemented correctly in accordance with the automated checks?
\end{description}

We will use the DISA Windows Server 2016 guide to answer \textbf{RQ1-4} and several CIS guides to answer \textbf{RQ5}.
For \textbf{RQ5}, we use CIS guides because, for them, we have the automatic checks and can assess a given system using their CIS-CAT tool.

\subsection{Degree of automation}
\label{sub:degreeofautomation}

To answer \textbf{RQ1}, \textbf{RQ2}, and \textbf{RQ3}, we examine the steps regarding the extraction of Windows policy automation using NLP and the verification of the found policy paths and values.
The results are depicted in Table~\ref{table:automation}.
From the 274 rules in the Windows Server 2016 guide, we can extract for 198 rules a possible policy setting with possible values.
Afterward, from the 198 possible configuration path/value pairs, 173 can be directly verified as valid configuration settings by the first verification step.
These 198 rules mean that for 63\% of the rules, we can extract both the policy path and the required value and verify that this value is valid for the particular policy path without any manual effort.
Thus, we could answer \textbf{RQ1}.
From the remaining 25 rules, for two rules, potential configuration settings and values have been extracted erroneously:
with our automation mechanisms, we could not automate these two rules.
We removed the erroneously created automations for theses two rules manually.\footnote{Tag: \href{https://github.com/tum-i22/disa-windows-server-2016/releases/tag/step-4a-fix-rules-which-have-been-imported-but-are-not-automatable}{step-4a-fix-rules-which-have-been-imported-but-are-not-automatable}}
Conversely, for four rules that we could automate, neither the policy path nor the required value was extracted.
In this case, we added the automation manually.\footnote{Tag: \href{https://github.com/tum-i22/disa-windows-server-2016/releases/tag/step-4b-fix-rules-which-have-not-been-imported-but-are-automatable}{step-4b-fix-rules-which-have-not-been-imported-but-are-automatable}}
Thus, the ratio of rules not added to the set of rules to automate, although they are automatable, lies at 1.5\%, whereas the ratio of rules which are not automatable and still extracted is 0.7\% regarding all rules.
For the remaining 23 rules that were extracted but could not be verified in the first round, we created the correct automation based on the extracted information enriched with the verification process's hints.\footnote{Tag: \href{https://github.com/tum-i22/disa-windows-server-2016/releases/tag/step-4c-fix}{step-4c-fix}}

If one sees the NLP based extraction process as a classifier with the classes \textit{automatable} and \textit{non-automatable}, the false-positive rate of this classifier is at 2.7\% and the false-negative rate at 2.0\%.
We had to adjust 27 rules manually.
Thus, for 90.1\% of all rules, respectively, 87\% of the automatable rules, no manual action was needed throughout the process.
In summary, these numbers answer \textbf{RQ2} and give strong evidence for the importance of our verification step because otherwise, these rules might be applied wrongly or not at all.

After the execution of the extraction and the verification step and the manual adjustments, we now have 200 rules which can be automated and have values that are verified to be valid for the given configuration decisions.
Therefore, the grade of automation we can achieve on the set of the 274 rules is at 73.0\%, respectively, at 96.6\% if we are only considering the 207 automatable rules (classified as automatable by DISA).
This number answers \textbf{RQ3}.
Thus, our approach reduces the number of rules which have to be checked or set manually on the system under test significantly.

\subsection{Time}

\begin{table}[t]
    \caption{Time needed to execute the single steps with all 200 automatable rules of the DISA Windows Server 2016 guide.}
    \label{table:time}
    \begin{tabular}{l r}
        \toprule
        Step                                    &   Time (s)    \\
        \midrule
        Knowledge Extraction from ADMX/L        &   81.59       \\
        Import into Scapolite                   &   8.02        \\
        NLP extraction of policy automations       &   16.93       \\
        Verification of policy automations         &   23.48       \\
        Export automations in JSON              &   13.90       \\
        Export automations in XLSX              &   14.03       \\
        Export policy backups from JSON            &   1.65        \\
        Check all rules for compliance          &   13.96       \\
        Implement automatable rules one-by-one  &   73.35       \\ \midrule
        \textbf{$\Sigma$}                       &   245.91      \\
        \bottomrule
    \end{tabular}
\end{table}

Table~\ref{table:time} shows time values for each of the automated steps.\footnote{
All the steps are conducted by running different commands from the command-line.
We ran every command 50 times and averaged the elapsed time to evaluate the speed of the single steps.
Configuration: 3.1 GHz Intel Core i7 with 16 GB RAM, Python 3.7.4.
The only steps implemented in PowerShell are the application viz. the check for compliance step as these were designed to be executed on the system under test, in our case, a Windows-based system, without installing any additional software. PowerShell Version 5.1.14393 was used.}
The short execution time per rule enables an application in CI approaches, which answers \textbf{RQ4} partially.

If we want to calculate the overall time, we also have to include the time it takes to correct the wrongly extracted automations.
According to Table~\ref{table:automation}, 25 rules were not verified the first time.
Because of the feedback included in the rule, we assume that it takes 10s to correct such a rule.
For the remaining four plus two rules, we assume that it takes at most 2min per rule to correct it.
These assumptions are also backed by the feedback of the users of our tools at Siemens.
Therefore, we end up with a total time of $ 245.91s + 25 * 10s + 6 * 120s = 1215.91s \approx 20min$ for all rules or $6s$ per rule.
Thus, \textbf{RQ4} could be answered, too.

\subsection{Correct Application}
\label{sub:correct_application}

In the last step of our evaluation, we want to answer whether our approach is applying the security-configuration guides correctly.
Incorrectly implemented rules can result from faults in the ADMX/L importer, the verification process, or the PowerShell library.
Here, our idea was that after applying a security-configuration guide to a system, the system should be configured as specified in the guide.
For this experiment, we use the standardized OVAL checks as ground truth.
Thus, we used guides which are Windows-related and for which we have automated checks.
Therefore, we used in this step 12 different security-configuration guides from the CIS, which are listed in Table~\ref{table:efficiency}, totaling over 2000 rules:
Four Windows-based OS's, six components of the Office package, and two browsers.

We conducted the evaluation as follows:
First, every security- configuration guide was automated through the same process, as explained in \S\ref{sec:genericapproach}.
Next, we set up a clean environment for every system.\footnote{
For the OS's, we have set up every system in a new VM by installing the OS directly from the latest ISO down-loadable from Microsoft.
As a VM provider, we used VirtualBox.
For the other components, e.g., Chrome or PowerPoint, they were directly installed on a clean Windows 10 instance.}
Additionally, we installed a SCAP-compliant scanner on the machines, i.e., the  CIS-CAT tool~\cite{ciscatpro}.
Next, we executed the checks in the \textit{clean} environment to compare the clean state with the hardened state to show that the implementation of guides makes the system more secure.
Afterwards, the guides are implemented using the automation generated as described in \S\ref{sec:genericapproach}.
Now the checks are rerun to test whether the implementation was correct.
The results are depicted in Table~\ref{table:efficiency}.
We also published the check reports before and after the implementation on GitHub~\cite{repoguides}.
Within this repository, one can find for every checked guide a \textit{before.html} and \textit{after.html} containing the result of the automatic check created using the CIS-CAT tool.

\begin{table*}
    \centering
    \caption{\# rules per guide compliant to the given guide before and after implementing guide automatically. Highest value of a column: dark gray, lowest: light gray.}
    \label{table:efficiency}

    \begin{tabular}{lrrrrrrrr}
\toprule
Guide               & \# of Rules & OVAL  & Before & \% & After & \%      & $ \Delta $ & $ \Delta $ \% \\
\midrule
Google Chrome for Windows               & 20                    & 20                 & \cellcolor{LG} 0  & \cellcolor{LG} 0 & 19                    & \cellcolor{LG} 95.0  & 19    & 95.0  \\
Internet Explorer 11    & 156                   & 136                & 1 & 0.7                 & 132                   & 97.1                 & 131   & 96.3  \\
Microsoft Office        & 53                    & 53                 & 2 & 3.8                  & 52                    & 98.1                 & 50    & 94.3  \\
Microsoft Access        & \cellcolor{LG} 9      & \cellcolor{LG} 9   & \cellcolor{LG} 0 & \cellcolor{LG} 0 & \cellcolor{LG} 9      & \cellcolor{HG} 100   & \cellcolor{LG} 9     & \cellcolor{HG} 100 \\
Microsoft Excel         & 34                    & 34                 & \cellcolor{LG} 0 & \cellcolor{LG} 0  & 34                    & \cellcolor{HG} 100   & 34    & \cellcolor{HG} 100 \\
Microsoft Outlook       & 75                    & 73                 & 3 & 4.1 & 72                    & 98.6                 & 69    & 94.5  \\
Microsoft PowerPoint    & 18                    & 18                 & 1 & 5.6 & 18                    & \cellcolor{HG} 100   & 17    & 94.4  \\
Microsoft Word          & 24                    & 24                 & \cellcolor{LG} 0  & \cellcolor{LG} 0 & 23                    & 95.8                 & 23    & 95.8  \\
Windows 7               & 390                   & 386                & 87 & 22.5               & 377                   & 97.7                 & 290   & 75.1  \\
Windows 8.1             & 429                   & 425                & \cellcolor{HG} 90 & 21.2 &  415                   & 97.6                 & 325   & 76.5  \\
Windows 10              & \cellcolor{HG} 505    & \cellcolor{HG} 502 & 85 & 16.9 & \cellcolor{HG} 489    & 97.4                 & \cellcolor{HG} 404   & 80.5  \\
Windows Server 2016     & 371                   & 334                & 88 & \cellcolor{HG} 26.3               & 325                   & 97.3                 & 237   & \cellcolor{LG} 71.0  \\
\midrule
$ \Sigma $              & 2084                  & 2014               & 357 & 17.7             & 1965                  & 97.6                 & 1608  & 79,8 \\
\bottomrule

    \end{tabular}

\end{table*}

Note that we only consider the rules which have OVAL checks for the calculation of the percentages.
We see that for the OSs between 16.9 and 26.3\% of the rules are already set up in a compliant way, whereas nearly no rule is pre-configured securely for the other components.
Nevertheless, even 26.3\% of already fulfilled rules of the OSs imply that the majority of the settings are configured in an insecure way on a clean system.
After applying the rules, the percentage of compliant rules is between 95 and 100\% for all guides.

%BG TODO I changed the text below because it seems to me that essentially these are all errors, either regarding the check or the specification of the implementation.

% Original Text
% ------------
%There are two significant reasons why the percentage of compliant rules is not at 100\% for every guide.
That we do not reach 100\% compliance
relative to the results of the CIS-CAT checker tool is due
to errors in the guides, some of them in the automated check, others in the descriptive text.
For example, some checks are overspecified, i.e., they expect more changes than actually occur when implementing the corresponding configuration setting:
the rule \textit{18.5.9.1} of the Windows 10 benchmark changes only a single registry entry, but the corresponding check refers to three different entries.
Also, some rules have automatic checks which test for wrong values. For example,
the check for the rule \textit{1.8.7.4} of the Word guide expects a different value (namely 0) than the value, which is set if the rule is implemented manually following the security-configuration guide.
Thus, we have in this rule precisely the difference of implementation and check we want to overcome with our approach.
Finally, there were some errors regarding the description of the implementation provided in the guides. For example, rule \textit{1.8.7.2.7} of the Word guide specifies that the setting should be enabled, although title and description suggest disabling the setting.
Another error in a guide actually is due to a  misspelling of the ADMX/L template file
provided by Microsoft. For example, rule \textit{1.13.2.1.5} of the Outlook guide specifies the value to be implement as \textit{When online always retrieve the CRL}, but our tool could not validate this value for this setting because of a misspelling in a template file. There, the value is written as \textit{When online always retr\textbf{ei}ve the CRL}.

All in all, we achieved compliance for 1965 rules (i.e., 97.6\%) after implementing the guides.
For the OSs, we have the highest absolute gain of compliant rules (between 237 and 404 rules), but in relative numbers, we are only gaining between 71\% and 80\%, whereas for the rest, we have a gain of over 90\%.
Please note that our approach can also implement the settings which were already compliant on a clean instance, but we have chosen this scenario because it seemed more relevant and natural.
The alternative would have been to create an instance in which every setting is configured to a non-compliant value.

\paragraph{Discussion}
\begin{listing}[b]
\begin{Verbatim}[commandchars=\\\{\}]
\PYG{g+gu}{\PYGZsh{}\PYGZsh{}} /implementations/0/description
Follow the below steps to disable \PYG{l+s+sb}{`Location Services`}:
\PYG{k}{1.} Tap \PYG{l+s+sb}{`Settings`} Gear Icon.
\PYG{k}{2.} Tap \PYG{l+s+sb}{`Security \PYGZam{} Location`}.
\PYG{k}{3.} Scroll to the \PYG{l+s+sb}{`Privacy`} section.
\PYG{k}{4.} Tap \PYG{l+s+sb}{`Location`}.
\PYG{k}{5.} Toggle to the \PYG{l+s+sb}{`OFF`} position.
\end{Verbatim}
\caption{Example of an implementation as part of a rule in an Android security-configuration guide.}
\label{lst:android_rule}
\end{listing}
\noindent
In \textbf{RQ1}, we asked for the percentage of rules for which we can automatically extract the implementation.
If our approach extracted the implementation only for a small fraction of rules, it would be useless in real-world applications.
Since we extracted for 63\% of all rules and 96\% of automatable rules an implementation, we can rule out this concern.

In \textbf{RQ2}, we looked for the percentage of false negative and false positives of our extraction process.
If these numbers were too high, the administrators would spend much time identifying them so that our approach would become pointless.
With 1\% and 2\% of the automatable rules wrongly classified, this is not the case.

In \textbf{RQ3}, we searched for the percentage of rules that we can automate after correcting the extraction process's errors.
If this number were too low, administrators would spend the same amount of time for implementing the remaining rules, and the gains of our approach would be small.
Our results show that we can automate 97\% of the automatable rules with our approach and dramatically reduce manual implementation.

In \textbf{RQ4}, we asked for the time taken to execute our approach.
If the steps were too time-consuming, it would be more efficient to do it manually, and our tooling would be unnecessary.
With 245.91s for the tools themselves and 1215.91s for the complete process, our approach is more efficient than the manual approach.

In \textbf{RQ5}, we searched for the percentage of rules which are correctly implemented according to the automatic checks.
If our approach implemented the rules wrongly, it would be useless.
With over 97\% of correctly implemented rules, our approach implements almost all rules correctly.

In summary, our evaluation showed that our approach is feasible and effective.

\section{Generalization and further work}
\label{sec:generalization}

\begin{listing}[b]
\begin{Verbatim}[commandchars=\\\{\}]
\PYG{n+nt}{ui\PYGZus{}name}\PYG{p}{:} \PYG{l+lScalar+lScalarPlain}{Location}
\PYG{n+nt}{namespace}\PYG{p}{:} \PYG{l+lScalar+lScalarPlain}{secure}
\PYG{n+nt}{name}\PYG{p}{:} \PYG{l+lScalar+lScalarPlain}{location\PYGZus{}providers\PYGZus{}allowed}
\PYG{n+nt}{value}\PYG{p}{:}
    \PYG{n+nt}{ON}\PYG{p}{:} \PYG{l+lScalar+lScalarPlain}{+network,+gps}
    \PYG{n+nt}{OFF}\PYG{p}{:} \PYG{l+lScalar+lScalarPlain}{\PYGZhy{}network,\PYGZhy{}gps}
\end{Verbatim}
\caption{Example of a definition for an Android-related setting.}
\label{lst:android_setting}
\end{listing}
\noindent
The main limitation of our extraction step is the fact that this extraction is only possible because of the highly schematic structure of the descriptions written by CIS and DISA.
If they modify their template for these descriptions, we will have to adjust this step entirely.
Thus, we hope that future guides will have the needed information in a machine-readable form.
A limitation of our implementation of Windows-related guides is the dependency on the \textit{LGPO.exe}.
If Microsoft decided to remove this tool for changing Windows system settings, we would have to replace core parts of the presented approach.

We admit that our approach is only an intermediate solution.
Instead of converting guides to executables by users or third parties, it would be more practical for publishers to attach machine-executable codes or links to them to the rules as they are doing it for automatic checking.
Nevertheless, as long as the publishers do not distribute the guides so that we can quickly and automatically implement them, we need tools like those we presented in this paper.

Our approach is tailored to Windows and its policies.
Thus, the approach cannot be ported to other platforms without significant adjustments.
Nevertheless, we are developing similar approaches for Linux OSs and Android in particular and try to achieve similar results there as well.

In Listing~\ref{lst:android_rule}, one can see the implementation of an Android-related rule.
It describes highly schematically the actions to implement.
Thus, the difficulty of the \textit{extraction} process as described in Figure~\ref{fig:implementation_process} is comparable to that for the Windows-related guides.

The \textit{verification} step is more difficult, because we do not have a similar definition of potential settings and the set of values they can have.
In Windows, we can extract this information from the ADMX/L files, but in Android, there are -- to our best knowledge -- no comparable files available.
To port our approach to Android, we created such definition files for several settings.
For the setting \textit{Location}, one would find an entry in this definition file as presented in Listing~\ref{lst:android_setting}.
With this information, we can verify that \textit{OFF} is a valid value for this setting.
Furthermore, we can use the information that we can translate \textit{OFF} to \textit{-network,-gps} for the \textit{transformation to a low-level automation}.
Finally, we can implement the rule on a given Android device via the Android Debug Bridge and the translated value.

Our work on Android just started, and there are many open questions:
How could be the syntax of an Android definition file?
How can we automatically create such a definition file?
Which settings can we automatically set, e.g., via the Debug Bridge, and which settings cannot be set or only if we have rooted the device?
How can we handle different Android versions and the fact that we can automatically configure a setting in one version, e.g., via the Debug Bridge, and in another version, it is no longer possible?

\begin{listing}[t]
\begin{Verbatim}[commandchars=\\\{\}]
\PYG{g+gu}{\PYGZsh{}\PYGZsh{}} /implementations/0/description
Set the following parameters in \PYG{l+s+sb}{`/etc/sysctl.conf`} or a
\PYG{l+s+sb}{`/etc/sysctl.d/*`} file:
\PYG{l+s}{```}
net.ipv6.conf.all.accept\PYGZus{}ra = 0
net.ipv6.conf.default.accept\PYGZus{}ra = 0
\PYG{l+s}{```}
Run the following commands to set the active kernel parameters:
\PYG{l+s}{```}
\PYGZsh{} sysctl \PYGZhy{}w net.ipv6.conf.all.accept\PYGZus{}ra=0
\PYGZsh{} sysctl \PYGZhy{}w net.ipv6.conf.default.accept\PYGZus{}ra=0
\PYGZsh{} sysctl \PYGZhy{}w net.ipv6.route.flush=1
\PYG{l+s}{```}
\end{Verbatim}
\caption{Example of an implementation in an Ubuntu Linux security-configuration guide.}
\label{lst:ubuntu_rule}
\end{listing}

For the automated implementation of general Linux guides, please have a look at Listing~\ref{lst:ubuntu_rule};
here, we have the implementation of a rule of a Ubuntu guide.
We can see that there is still a schema of how the implementation is described.
Nevertheless, it is more complicated. In this example, there are two different steps, one concerning the modification of a file, the other the execution of shell commands. Hence,
in addition to extracting the code-snippets, we have to derive the semantics of \textit{set file content to} and \textit{run} as well.
If we wanted to \textit{verify} that the code snippets are valid, we would have to know the syntax of the specific configuration file and the semantics, e.g., if \textit{net.ipv6.conf.all.accept\_ra = 0} is a correct line in this file.
Furthermore, we would have to know the legal parameters of the program called in the second snippet.

In our future research, we will try to extract this information, e.g., from the source code, the documentation, or sample configuration file to create definition files for the most common commands and configuration files.
% BG: Saving space below
%Whereas the \textit{extraction} and \textit{verification} is more difficult for the Linux-based systems, the automatic implementation should be straightforward as we can use existing solutions like Ansible.
%Thus, we would only have to translate the extracted information into a format of one of these tools.
% BG: I changed the outlook below a bit:
% original text
% --------------
%In summary, we think that it might not be easy but that it is doable to port our approach from Windows to Linux-based systems.
%We will present our results on these topics in future articles.
In summary, we think that our approach can be ported from Windows to Linux-based systems, but whether a comparably high percentage of rules from which automations can be extracted can be reached remains to be seen.

%We will present our results on these topics in future articles

In the future, work is necessary to provide the foundations that make security automation easier.
The main factor that made our approach possible was that Microsoft provides machine-readable information about configuration options for their systems in the form of ADMX/L files.
It follows that vendors should support security automation by providing machine-readable information about security-configuration options and their implementation.
%BG: Saving space below
%Work on security automation, e.g., by SACM and NIST regarding SCAP v2, should promote the creation of appropriate standards.

%% file: contents/09_related_work.tex
\section{Related Work}
\label{sec:related_work}

Many studies have been conducted in the field of misconfiguration, e.g. ~\cite{Dietrich:2018:ISO:3243734.3243794,Tang:2015:HCM:2815400.2815401,Continella:2018:THB:3274694.3274736,Jha2017,Xu:2015:SAT:2775083.2791577}.
Especially the study of Dietrich et al.~\cite{Dietrich:2018:ISO:3243734.3243794} is relevant for our research.
Their study provides strong evidence that security misconfigurations are more common than usually assumed.
This emphasizes how important and yet underestimated this field of research currently is.
Furthermore, they have identified the lack of knowledge and experience as core factors for security misconfiguration and argue that we need more automation in the whole process to make systems more secure.
By using security-configuration guides, we want to tackle the first problem, with our automated implementation the second.

Additionally, many researchers explored how to detect and how to avoid misconfigurations~\cite{Rahman:2019:SSS:3339505.3339528,Santolucito:2017:SCF:3152284.3133888,Keller2008,Su:2007:AIC:1323293.1294284}.
Rahman et al.~\cite{Rahman:2019:SSS:3339505.3339528} analyzed thousands of Infrastructure-as-a-Code (IaaC) scripts to identify insecure configurations and security smells.
They used these smells to create a linter for creating more secure IaaC scripts.
Although their linter is comparable to the hints we give to the administrators, we are targeting different problems.
Where they are extracting knowledge from the IaaC scripts on how to configure systems securely, we already have this information and have to apply it.
Furthermore, as discussed before, we think that IaaC scripts are not sufficient to specify security-configuration guides.
Similar work was done by Santolucito et al.~\cite{Santolucito:2017:SCF:3152284.3133888}.
Their framework \textit{ConfigV} aims at similar problems as our verification step.
In contrast to them, we cannot learn secure configurations.
Instead, these are defined in the guides, and the constraints do not have to be learned but can be extracted from the ADMX/L files.
Similarly, SPEX, developed by Xu et al.~\cite{Xu:2013:BUM:2517349.2522727} is not applicable in our case, as we do not have the source code of the programs we want to configure.

Raab et al.~\cite{DBLP:conf/oss/RaabB17,DBLP:conf/aosd/Raab16,DBLP:conf/caise/Raab13} created the \textit{Elektra} framework to validate the access to configuration values to detect misconfigurations as soon as possible.
We tried to achieve the same with our a-priori verification process.
In their study~\cite{DBLP:conf/oss/RaabB17}, they investigated how free/libre and open-source software (FLOSS) can be configured and the problem of validating configurations for it.
One finding is that presently, configuration validation is encoded in a way unusable for external validation or introspection tools.
Although Windows is not a FLOSS, we encountered the same problem.
This is why we had to implement our verification mechanism instead of simply using an existing tool.
Furthermore, Elektra is tailored towards developers who create new software, not for administrators of existing software and it cannot handle the Windows policy settings we have to change according to the guides.
Thus, we could not apply Elektra.

A similar approach to Elektra was developed by Xu et al.~\cite{Xu2016} with the same problems so we could also not use it in our case.
In their study~\cite{Xu:2015:HYG:2786805.2786852}, they have shown how the growing complexity of the configuration of systems is overwhelming users and systems administrators.
They did not investigate Windows systems, yet many of their findings apply to our domain, too. For instance, users have tremendous difficulties because they do not know which parameters to set and that this induces up to 50\% of the configuration errors.
This supports the claim that we need security-configuration guides created by experts, to be used by system administrators.

Wang et al.~\cite{Wang:2008:TAR:1455770.1455802} present an approach at automatic reverse engineering of an application’s access-control configurations.
Although the application domain is similar to our context, we could not apply their work for our need as we do not have the source code of the programs we want to configure securely.

There also is a lot of research in the field of extracting important parameters or configuration values from human-readable documents~\cite{Yang2018,Wong:2015:DDS:2818754.2818831,Pandita:2012:IMS:2337223.2337319,Tan:2007:IBB:1323293.1294276,Zhong2011,Rabkin2011,Sayagh2020,Jin:2014:PGR:2642937.2643009}.
Yang et al.~\cite{Yang2018} present an approach to automatically extract web API specifications from the documentation of a software similar to the extraction of our configuration values from the security-configuration guidelines.
However, the fact that our documents do not contain as many links made this approach unfeasible in our case.
Using NLP, Wong et al.~\cite{Wong:2015:DDS:2818754.2818831} developed an approach at extracting information from program documentation to improve automated testing.
They use grammar rules to identify relevant comments and extract constraints from them.
In our case, the security-configuration guides describe concepts from a higher level then program documentation.
Furthermore, we do not need to extract the constraints from the security-configuration guide.
Thus, this approach was also not applicable in our context.

Closest to our work regarding our aims of providing rule-by-rule implementation is the OpenScap project~\cite{openscap,preisler:haicman:UsenixLisa:2017}.
OpenScap maintains its security-configuration guides for various Linux systems in a git repository, where each rule is represented by one file;
usually, the file holds references to other files containing artifacts for automated implementation and check.
However, we cannot use OpenScap.
First, OpenScap cannot implement the rules from the Windows-based guides.
Second, if OpenScap could implement them, we would first have to add the scripts manually to the guides of CIS or DISA.
We think that the guides in the context of OpenScap are one step ahead of Windows guides published by CIS or DISA because of the connection between implementation and checking.
In the future, we hope that the publishers distribute their Windows guides similarly to OpenScap in a form that is as easily implementable as checkable.
We consider our approach an intermediate solution to bring the automatic implementation of Windows-based guides to a comparable level as long as this is not the case.

Ongoing activities regarding further improvements of automating security as carried out by the IETF SACM (\emph{Security Automation and Continuous Monitoring}) work group~\cite{ietf-sacm-terminology-16,ietf-sacm-arch-00,rfc8248}
as well as a first indication of the direction work towards SCAP version 2 as outlined in a transition document~\cite{scap_transitioning}, have a clear focus on checking security-configuration settings and disregard their implementation---which is precisely the gap we want to close in this work.

To sum up the related work:
some approaches use NLP to extract settings from the documentation or the source code of a program, but to our best knowledge, no approach extracted the settings from security-configuration guides.
Furthermore, some approaches like Elektra help to improve the configuration of newly developed software, but we cannot use them to configure existing and closed-source Windows systems.
We can automatically implement the guides of some Linux variants with the OpenScap approach if the publishers distribute them with the scripts necessary for OpenScap.
However, we cannot use OpenScap to implement existing Windows-based guides automatically.
Thus, we tackled these gaps in the literature and put the developed components together to demonstrate that our proposed approach and our PoC implementation are achieving promising results.

%% file: contents/11_conclusion.tex
\section{Conclusion}
\label{sec:conclusion}

The complexity of contemporary systems renders their configuration increasingly difficult.
This leads to vulnerabilities attackers can exploit to attack the systems.
For a single organization, it is impossible to know all the configurations to make a specific system secure.
Many organizations use public security-configuration guides to overcome the lack of knowledge;
while many of these guides support automated compliance checking, they do not provide support for automated implementation.
%but in general, we cannot automatically implement them.

In this paper, we demonstrated an approach that can automatically implement Windows security-configuration guides with minimal manual effort.
%automatic implementation of these guides almost entirely in the Windows context.
%This significantly eases the hardening work of system administrators and also enables, for the first time, private persons to harden their Windows-based system with an acceptable effort.
Our contribution further encompasses a
%consists of our approach at automatically implementing existing Windows-related security-configuration guides;
proof-of-concept implementation, 
a step-by-step documentation of the process, 
and the evaluation of our approach using existing guides.

Our evaluation has shown that we can automate 83\% of the rules without any manual effort using our NLP extraction.
Furthermore, our extensive benchmark with 12 different guides and over 2014 rules with automatic checks showed that the implementation of our approach can implement at least 97\% of the rules correctly.

With our approach and the results of its evaluation, we believe we can furthermore contribute as follows:
Firstly, we have demonstrated how organizations
that rely on publicly available security-configuration guides can be aided in reducing effort as well as reducing
errors in the implementation of these guides.
Secondly, we have shown how machine-readable information supporting automated implementation for Windows systems can be represented and included in SCAP guides. We hope that our results encourage publishers of security guides to support better the automated implementation of their guides by enriching them with such information, for Windows as well for other target systems. The design of SCAP v2 has already started~\cite{scap_transitioning}:
Our work offers timely and relevant input for the further development of SCAP towards a standard that meets the requirements of both publishers and consumers of machine-readable security-configuration guides.

Thirdly, our research underlines the need for machine-readable specifications of (security) configuration settings: standardization and support of a format for this purpose by vendors would significantly aide in all tasks concerned with configuring systems securely. 
%we hope that more vendors will provide specification documents such as Microsoft's ADMX/L files to enable the verification of automation.
%This could improve enormously the creation process of security-configuration guides.

%% file: contents/12_acknowledgements.tex
%\section{Acknowledgments}

% \begin{acks}
% \end{acks}

%% file: contents/13_availability.tex
%\section{Availability}
%\label{sec:availability}
%
%NB: An example of how an existing security-configuration guide in the SCAP format can be transformed and automated is, as aforementioned, publicly available on GitHub~\cite{steprepository}.
We plan to release substantial parts of our Python code-base.
%for creating and manipulating Scapolite-based data as open source as well as the PowerShell library for implementation and check of security-configuration guides on Windows-based systems.
% At the time of writing, however, it is unclear whether this will be possible in time for the USENIX Security Symposium.